\documentclass[12pt,preprint]{aastex}

\shortauthors{Ott et al.}
\shorttitle{Hot Gas in NGC\,3077}

\begin{document}

\title{Chandra Observations of Expanding Shells in the Dwarf Starburst Galaxy NGC\,3077}

\author{J\"urgen Ott\altaffilmark{1}}
\affil{Radioastronomisches Institut}
\affil{Universit\"at Bonn\\ Auf dem H\"ugel 71\\53121 Bonn, Germany}
\email{jott@astro.uni-bonn.de}
\vspace{0.5cm}

\author{Crystal L. Martin\altaffilmark{2}}
\affil{University of California, Santa Barbara}
\affil{Department of Physics\\ Santa Barbara, CA 93106, USA}
\email{cmartin@physics.ucsb.edu}

\and

\author{Fabian Walter\altaffilmark{3}} 
\affil{National Radio Astronomy Observatory}
\affil{P.O. Box O}
\affil{Socorro, NM 87801, USA}
\email{fwalter@nrao.edu}

\altaffiltext{1}{Current Address: Bolton Fellow, CSIRO Australia Telescope National Facility, Cnr Vimiera \& Pembroke Roads, Marsfield NSW 2122, Australia; email: Juergen.Ott@atnf.csiro.au} 
\altaffiltext{2}{Packard Fellow}
\altaffiltext{3}{Janksy Fellow}

\begin{abstract}
Deep Chandra observations (53\,ks, ACIS--S3) of NGC\,3077, a starburst
dwarf galaxy in the M\,81 triplet, resolve the X--ray emission from
several supershells. The emission is brightest in the cavities defined
by expanding shells detected previously in H$\alpha$ emission (Martin
1998). Thermal emission models fitted to the data imply temperatures
ranging from $\sim 1.3$ to $4.9\times 10^{6}$\,K and indicate that the
strongest absorption is coincident with the densest clouds traced by
CO emission. The fitted emission measures give pressures of
$P/k\approx 10^{5-6} \,\xi^{-0.5}\,f_{v}^{-0.5}$\,K\,cm$^{-3}$ ($\xi$:
metallicity of the hot gas in solar units, $f_{v}$: volume filling
factor). Despite these high pressures, the radial density profile of
the hot gas is not as steep as that expected in a freely expanding
wind (e.g., as seen in the neighboring starburst galaxy M\,82)
implying that the hot gas is still confined by the H$\alpha$ shells.
The chaotic dynamical state of NGC\,3077 undermines reliable estimates
of the escape velocity. The more relevant quantity for the ultimate
fate of the outflow is probably the gas density in the rich intragroup
medium. Based on the \ion{H}{1} distribution of NGC\,3077 and a
connected tidal tail we argue that the wind has the potential to leave
the gravitational well of NGC\,3077 to the north but not to the south.
The total $0.3-6.0$\,keV X--ray luminosity is $\sim 2-5\times
10^{39}$\,erg~s$^{-1}$ (depending on the selected thermal plasma
model). Most ($\sim 85$\%) of the X--ray luminosity in NGC\,3077 comes
from the hot interstellar gas; the remainder comes from six X--ray
point sources. In spite of previous claims to the contrary, we do not
find X--ray emission originating from the prominent tidal tail near
NGC\,3077.

\end{abstract}


\section{Introduction}
\label{sec:intro}

In lower mass galaxies, violent star formation activity is predicted
to drive interstellar gas into either the dwarf galaxy's halo or the
intergalactic medium \citep[IGM;][]{dek86,mac99,sil01}. Cold dark
matter cosmologies predict that dwarf galaxies began forming before
more massive galaxies could be assembled. The fate of the dwarf galaxy
winds may therefore have influenced the formation of the larger
galaxies. It is not well understood, however, whether the properties
of the surrounding intragroup gas or the gravitational potential
dictate the fate of such galactic winds. Although intergalactic gas
densities are much lower at the present epoch, detailed examination of
the wind/group interaction in the local universe should improve our
understanding of it at all cosmic epochs.

It has long been known through deep observations in the H$\alpha$ line
that galactic outflows exist, but relatively little is known about the
hot phase of the gas ($T\sim10^{6-7}$\,K) which actually drives the
outflow. Early studies with EINSTEIN and ROSAT of extremely luminous
starburst galaxies revealed the presence of hot winds, sometimes
reaching distances above the plane of several kpc (e.g., M\,82:
\citealt{str97} in the following S97; \citealt{leh99}, NGC\,253:
\citealt{str00b,str02,pie01}). Measurements of the temperature and
density in the hot winds were critical for estimating their power
requirements and establishing the mechanical power from supernovae in
the starburst as the energy source \citep[see, e.g, the reviews
of][]{hec93,hec02}. The latest generation of X--ray observatories,
Chandra and XMM--Newton, allow similar measurements of smaller,
fainter galaxies due to their larger collecting areas. The angular
resolution of Chandra, $\sim 1\arcsec$, also allows a) the direct
identification of discrete X-ray sources and b) spectral imaging of
extended gas structures on spatial scales attained in optical and
radio maps. Such detailed information turns out to be crucial for
establishing a physical model of the rather complex integrated X-ray
spectrum.

In this paper we present Chandra X--ray observations of the dwarf
starburst galaxy NGC\,3077. We chose this particular dwarf galaxy for
our study because NGC\,3077 shows prominent expanding H$\alpha$
supershells \citep[][in the following M98]{mar98} and is a member of
the gas-rich M81 group. Interaction with M\,81 and M\,82 is believed
to have triggered starburst activity in both M\,82 and NGC\,3077.
Tidal tails in the M\,81 triplet allow to investigate environmental
effects on the development of superwinds. Various studies of
NGC\,3077 have been published in the past (e.g., optical studies:
\citealt{bar74}, atomic and molecular gas: \citealt{wal02}, in the 
following W02; \citealt{mei01}). Previous ROSAT X--ray observations
indicating the presence of diffuse X--ray emission in NGC\,3077 were
presented by \citet{bi94}. The interpretation of these observations,
however, are complicated due to point source confusion.

We adopt a distance to NGC\,3077 of 3.6\,Mpc \citep[as obtained for
M\,81,][]{fre94}, consistent with more recent studies
\citep[cf.][]{sak01,kar02}. In Sect.\,\ref{sec:obs} we describe the
Chandra observations, the data reduction, and the results. A
discussion is presented in Sect.\,\ref{sec:discuss} followed by
conclusions and a summary in Sect.\,\ref{sec:summary}.

\section{Observations, Data Analysis, and Results}
\label{sec:obs}

We observed NGC\,3077 with the Chandra X--ray Observatory on 2001
March 07 and 08 for 53.4\,ks with the back--illuminated Advanced CCD
Imaging Spectrometer (ACIS) S3 CCD chip (pixel size: $0\farcs49 \times
0\farcs49$, energy resolution $\sim 120$\,eV at 1\,keV; observation
ID: 2076, Seq.No. 600210). ACIS was at a temperature of
$-120^\circ$\,C during these observations. The satellite telemetry was
processed at the Chandra X-ray Center (CXC) with the Standard Data
Processing (SDP) system to correct for the motion of the satellite and
to apply cosmic ray rejection. Since the data were processed before
2001 September 13, we applied a new gain map to accommodate the latest
Order--sorting/Integrated Probability Tables (OSIP). The data products
were then analyzed with the CXC Chandra Interactive Analysis of
Observations (CIAO) software version 2.2. CIAO Data Model tools were
used for data manipulation, and Sherpa v\,2.2 for the fitting of the
ACIS spectra. CALDB v\,2.9 of the ACIS CCD calibration files were
applied throughout the analysis. We constructed a lightcurve by
binning the dataset to a time resolution of 200\,s. No significant
flaring of the background was observed. Bad pixels of all images were
screened out by applying the bad pixel mask provided by the CXC. We
checked the Chandra pointing during our observations by verifying the
position of the bright double star HD\,86677. Its position falls
within 1\arcsec\ of the measured coordinates listed in the HIPPARCOS
Catalog \citep{per97}.

\subsection{Global Properties}

\label{sec:various_x}

We defined four broad energy bands for the analysis of the X--ray
emission of NGC\,3077 in an attempt to separate strong oxygen lines
(soft band: 0.3\,keV$\leqslant E \leqslant$0.7\,keV) from major
iron--L line complexes (medium band: 0.7\,keV$\leqslant E \leqslant$
1.1\,keV) of a hot thermal plasma. Line emission does not play an
important role for higher energies. The definition of the third band
(hard band: 1.1\,keV$\leqslant E \leqslant$ 6.0\,keV) therefore
provides a good representation of the continuum emission. As
photoelectric absorption (cross section $\sigma$) is a function of
energy $E$ ($\sigma \propto E^{-3}$), it is strongest in the soft
band. The hard band, in contrast, is a better tracer for
temperature. We also defined a total band (0.3\,keV$\leqslant E
\leqslant$ 6.0\,keV) to discuss the global characteristics of
NGC\,3077. The resulting images were adaptively smoothed with the
CIAO task {\it csmooth}, using a fast Fourier algorithm and a minimum
and maximum significance S/N level of 2 and 3, respectively. The
levels were chosen this way in order to give the best representation
of features which are visible in the raw data set. Adaptively smoothed
images, however, were not used for any quantitative analysis.

Exposure maps were created for the soft, the medium, and the hard band
based on their central energies. Since the intensity of the diffuse
emission peaks at about 1\,keV, we decided to use the exposure map of
the medium band for the total band, too. The smoothing kernel of the
adaptively smoothed images changes size for each pixel, and exposure
map correction must account for this effect. To do so, we smoothed the
exposure maps by using the smoothing scales which were previously
computed for the adaptively smoothed images. All images were finally
corrected by dividing by the appropriate exposure map.

The distribution of the X--ray emission in the different bands are
shown in Fig.\,\ref{fig:bands} -- Fig.\,\ref{fig:color}\,a is a three
color composite of the same images. Point sources contribute
$\sim$27\% to the total count rate of NGC\,3077 (see
Sect.\,\ref{sec:detect}) and are labeled in Fig.\,\ref{fig:bands}; the
remaining 63\% is diffuse emission. The diffuse X--ray emission of
NGC\,3077 has a size of about 1\arcmin\ (corresponding to 1\,kpc).
The soft emission is slightly more extended than the medium and hard
emission (see discussion in Sect.\,\ref{sec:diffuse_reduc}),
especially to the east of NGC\,3077 (see Fig.\,\ref{fig:bands}).

\placefigure{fig:bands}
\placefigure{fig:color}

We constructed two hardness ratios from the bands defined above:
\begin{eqnarray}
{\rm HR1}&=&{\rm (Soft}-{\rm Medium}-{\rm Hard)/(Soft+Medium+Hard)}\label{eq:hr1}\\
{\rm HR2}&=&{\rm (Soft+Medium}-{\rm Hard)/(Soft+Medium+Hard)}
\label{eq:hr2}
\end{eqnarray}

HR1 is more sensitive to photoelectric absorption than HR2, while HR2
can be used to trace temperature variations. The advantage of defining
the hardness ratios via three bands compared to a more classical two
band definition is that the values are still defined for all sources
even if virtually all emission emerges from within one band. 

In Fig.\,\ref{fig:hr_images} we present the hardness ratio images of
NGC\,3077 based on the adaptively smoothed broad band images. Due to
the non--linearity of the {\it csmooth} algorithm, deviations from the
real distributions may appear. Still, this representation gives a good
overview on the global characteristics of the hardness ratios within
NGC\,3077. We find that the point source S4 is the softest source and
S2 and S3 are the hardest. The hardest diffuse X--ray emission is
found near the center of NGC\,3077, close to S1, S2, and S3. To the
north and west of NGC\,3077, the X--ray emission is brighter in the
medium band as compared to the soft band.

\placefigure{fig:hr_images}
\placefigure{fig:hard_hi}

Photoelectric absorption of X--ray emission leads to a hardening of
the spectrum as it is most effective at low photon energies ($\propto
E^{-3}$). One would expect that this effect happens where cold gas is
located in front of the X--ray emission. To test this, we compare the
HR1 image to the \ion{H}{1} and the CO distributions of NGC\,3077 in
Fig.\,\ref{fig:hard_hi} (data taken from W02; see also
Fig.\,\ref{fig:color}c). The dense molecular clouds are located where
the center of NGC\,3077 exhibits the hardest diffuse X--ray emission.
Moreover, the \ion{H}{1} emission follows nicely the HR1=$-$0.3
contour. In summary, the X--ray absorption features seen in NGC\,3077
can potentially be explained by the presence of atomic and molecular
gas.

\subsection{Discrete Sources}
\label{sec:pointsources_reduc}

\subsubsection{Source Detection}
\label{sec:detect}

In order to separate point sources from diffuse emission, we applied
the wavelet source detection algorithm {\it wavdetect} to the
spatially unbinned, total band image. The point spread function (PSF)
degrades at large off--axis angles and we accounted for this effect by
using $0\farcs5$, $1\farcs0$, and $2\farcs0$ wavelet radii. Using this
method, 42 sources were detected on the ACIS--S3 chip; four of the
detections are located within the diffuse X--ray and H$\alpha$
emission of NGC\,3077. To cross--check this result, we ran the
alternative sliding cell {\it celldetect} algorithm on the same
dataset. Regarding the varying shape of the PSF, the signal to noise
(S/N) thresholds were chosen to be 1.9, 2.1, and 3.2, where the first
threshold was applied within the inner 1\arcmin\ radius centered on
the aimpoint. The second and third thresholds were used within radii
$r$ of $1\arcmin < r < 2\farcm5$ and $2\farcm5 < r < 3\farcm5$. Beyond
3\farcm5, the PSF degrades substantially and detections were generally
disregarded \citep[for a justification see][]{mar02}. This led to a
total detection of 28 sources, where again four belong to
NGC\,3077. No additional sources were detected by {\it
celldetect}. Sources with low S/N were more likely to be detected by
{\it wavdetect} than by {\it celldetect}. In Table\,\ref{tab:field} we
list the detections within the ACIS--S3 field of view (excluding those
located within the optical extent of NGC\,3077; they are listed
separately in Table\,\ref{tab:pointsources}). The source numbers
(column 1) are the same as in Fig.\,\ref{fig:field}. Column (2) lists
the detection algorithm that detected the source (w: {\it wavdetect},
c: {\it celldetect}, x: the distance to the aimpoint is $>3\farcm5$;
{\it celldetect} was not applied). The net source counts in the
different bands are given in columns (5) to (8) and the hardness
ratios HR1 and HR2 in columns (9) and (10). A star at the source
number denotes an optical counterpart from an R$_c$--band image within
a search radius of 10\arcsec (R$_c$ and B--band images obtained by us
at the Calar Alto 2.2\,m telescope with integration times of 600\,s
and 900\,s, respectively). This radius was chosen to account for
seeing and positional uncertainties of the ground based data as well
as for the extents of the tentative optical counterparts.

\placefigure{fig:field}

\placetable{tab:field}

\subsubsection{Point sources in NGC\,3077}
\label{sec:3077point}

As mentioned above, both source detection algorithms detected four
discrete sources within the diffuse X--ray and H$\alpha$ emission of
NGC\,3077 which supposedly belong to this galaxy. From a visual
inspection of the total band image, we decided to add a source close
to the center of NGC\,3077 to our list (S5, see
Fig.\,\ref{fig:bands}). One detection exhibits two intensity peaks
separated by 1\arcsec\ which is about the size of the ACIS--S3 PSF at
1\,keV (FWHM[PSF] $\approx 1\farcs1$). We decided to split this
detection into two sources (S2 and S3) for separate analysis. To check
whether or not the discrete sources are indeed point--like, we
constructed a more accurate PSF for each source position, given the
intensity--weighted mean energy of the corresponding source spectrum.
Except for S5 which appears to be slightly extended (deconvolved size:
$1\farcs1 \times 1\farcs3$), all other sources are indeed point
sources.

Spectra of the point sources were extracted from apertures which were
chosen to be three times larger than the corresponding PSF, except for
S2 and S3; in these cases the spectra were extracted from rectangular
boxes $3\farcs0 \times 2\farcs3$ in size in an attempt to separate
both objects (see Fig.\,\ref{fig:bands}). The corresponding
Redistribution Matrix Files (RMFs) and Auxiliary Response Files (ARFs)
were subsequently created for each source position on the CCD to
calibrate the corresponding spectra. The spectral channels were not
binned for any spectral fitting analysis in order to preserve the
maximum of statistical information. One should keep in mind, however,
that the spectral resolution of ACIS--S3 is about 120\,eV at 1\,keV
which corresponds to about eight pulse invariant (PI) energy channels.

Each spectrum is composed of the real source spectrum and an
energy--dependent offset which comprises instrumental and
extragalactic backgrounds as well as diffuse X--ray emission from
NGC\,3077 itself. We checked the intensity of these offsets by
extracting source counts in regions with an X--ray contribution
similar to that expected at the source positions. This was obtained
within apertures of the same size as those used for the extraction of
the point source spectra. As only 1--3 counts of background were
detected within each aperture, we decided to neglect the background
contamination of the X--ray spectra of the discrete point source
population.

The final point source spectra can be divided into three types: a)
flat spectrum sources with counts all across the ACIS energy range
(sources S2 and S3), b) source spectra which peak at energies
$\sim$0.8--1.2\,keV (S1, S5, and S6), and c) a spectrum (S4), which
peaks at $\sim$0.6\,keV and virtually shows no emission above
0.8\,keV.

The spectra were modeled within the energy range of 0.3\,keV$\leqslant
E \leqslant$ 8.0\,keV (neglecting data with bad quality flags). We
fit three different models to each spectrum: a Raymond--Smith
collisional thermal plasma \citep[RS,][and updates\footnote{see the
 XSPEC manual:
 http://heasarc.gsfc.nasa.gov/docs/xanadu/xspec/manual/manual.html }\ 
]{ray77}, a power law (PL) and a black body (BB) model. All source
models were coupled with the photoelectric absorption model from
\citet{bal92}. Following \citet{hec80} and \citet{mar97}, we assumed
solar metallicity for NGC\,3077. This value is in good agreement with
the global optical spectrum of NGC\,3077 \citep{ken92} as applied to
the oxygen abundance estimation method recently developed by
\citet{kew02}. The solar element mixture, which is implemented in the
Sherpa code was adopted from \citet{and89}. The metallicity of
NGC\,3077 is similar to that of our Galaxy. For this reason, it is
sufficient to model the photoelectric absorption (due to the Galactic
interstellar medium [ISM] as well as to material within NGC\,3077) by
a single component, with the restriction that the derived column
densities must exceed the Galactic value. From the Leiden/Dwingeloo
Survey \citep{har97} we measure a Galactic \ion{H}{1} column density
of $4\times 10^{20}$\,cm$^{-2}$ at the position of NGC\,3077.

We used the Cash maximum likelihood statistics for best fit
estimation, which is more accurate than the $\chi^{2}$ statistics in
the case of low number statistics. The optimization method was a Monte
Carlo algorithm with 64 random starting values within a physically
sensible parameter space. Note that a lower number of starting values
provided unstable results. For each of these initial parameters, the
single--shot Powell optimization method locally maximized the
likelihood ({\it Monte--Powell} within Sherpa). The fits with the
supposedly global maximum likelihood were finally confirmed by a
Levenberg--Marquardt optimization method algorithm. The best fit
parameters for each source and model are shown in
Table\,\ref{tab:pointsources}, where $N_{H}$ is the absorbing column
density, $T$ the temperature, $\gamma$ the power law photon index,
$F_{X}^{abs}$ the absorbed flux, $F_{X}$ the unabsorbed flux, and
$L_X$ the unabsorbed X--ray luminosity. Unfortunately, the surfaces of
constant likelihood in the parameter spaces especially of absorbed
thermal plasma models are not well--behaved, i.e., they are not
multi--dimensional paraboloids. The use of any standard error
estimation algorithm like Sherpa's {\it uncertainty}, or {\it
 covariance} will therefore inevitably fail. For this reason, errors
of the individual fits were obtained by using the ``frequentist's
method'' \citep[see, e.g.,][]{fre01}. For each set of best fit
parameters we simulated 500 spectra with the Poisson noise
distribution of the observations. Each of these simulations were
re--fitted using the best fit values as initial values for a Powell
optimization method. The resulting values for all parameters were used
to construct a sampling statistic and the inner 68\% of the
distribution obtained are considered to be good estimates for the
1\,$\sigma$ errors. Errors for fluxes and luminosities were calculated
on the basis of extreme parameter values. The most plausible best fits
are displayed in Figs.\,\ref{fig:p1} and \ref{fig:p23456} (see
Sect.\,\ref{sec:discuss_pointsources} for a discussion) as well as the
corresponding confidence regions in the $N_{H}$--$T/\gamma$ planes.
Unfortunately, the limited S/N does not allow to give preference to
any specific model based on a determination of a goodness--of--fit
indicator.

\placefigure{fig:p1}

\placefigure{fig:p23456}

\placetable{tab:pointsources}

We checked all discrete point sources for signs of variability. To do
so, we binned the data into 60\,s, 200\,s, 600\,s and 1000\,s time
bins. A visual inspection of the lightcurves did not reveal
variability for any of the discrete sources at a level significantly
exceeding the errors of the observation.

Finally, we compared the X--ray point source population of NGC\,3077
with optical B and a R$_c$ band images (see
Sect.\,\ref{sec:detect}). Optical counterparts were not detected. An
additional SIMBAD query did not reveal known sources coinciding with
any of the discrete X--ray sources.

\subsection{Diffuse Emission}
\label{sec:diffuse_reduc}

To obtain broad band images of diffuse emission, the discrete sources
discussed in Sect.\,\ref{sec:3077point} were subtracted from the data
over an area set by three times the corresponding PSF. The blanked
areas were subsequently refilled by a Poisson distribution
interpolated from nearby regions within the area of diffuse emission
(task {\it dmfilth} in CIAO). The resulting images were then used to
construct azimuthally averaged surface brightness profiles of each
band, taking the infrared H--band peak as the center
\citep[$\alpha_{J2000} = 10^h03^m19^s$,
$\delta_{J2000}=68\degr44\arcmin02\arcsec$ derived from HST NICMOS
data,][see also Fig.\,\ref{fig:bands} and \ref{fig:color}b]{boe99}.
Logarithmic representations of these profiles are displayed in
Fig.\,\ref{fig:surflog}. All bands are consistent with an exponential
decline which is supported by corresponding fits to the data. Most of
the X--ray photons are emitted in the medium band. The scalelength $h$
of the surface brightness profile of this band ($200\pm 9$\,pc) is
thus similar to the one of the total band ($h=186\pm 10$\,pc). In
contrast, the soft energy band decreases more slowly ($h=296\pm
49$\,pc) and the hard band more rapidly ($h=135\pm 21$\,pc). The
scalelengths quantitatively corroborate the morphological picture
discussed in Sect.\,\ref{sec:various_x}.

\placefigure{fig:surflog}

With the help of the azimuthally averaged surface brightness profile
of the total band, we were able to construct a radial volume density
profile. For simplicity we assumed a spherical geometry of the hot
coronal gas and decomposed the morphology into three--dimensional
onion--like shells coincident with the two--dimensional annuli defined
for the surface brightness profiles (for an example see the inset of
Fig\,\ref{fig:density}). The emission measure $EM$ detected in each
annulus $k$ is the sum of the squared densities $n$ within the
corresponding shells $i$ multiplied by their lines of sight $l$; the
general expression being

\begin{equation}
EM(k)=\sum_{i=k}^{M}\,(\xi f_{v})^{-1}\,n(i)^{2}\,\, l(k)_{i}.
\end{equation}

$M$ denotes the outermost annulus, where the density can be easily
calculated via $n(M)=\sqrt{EM(M)/l(M)_{M}}$. We further assume that
the diffuse emission is emitted by a volume filling ($f_{v}=1$) hot
collisional thermal RS plasma, with a temperature of $2\times
10^{6}$\,K (solar metallicity $\xi=1$ and solar element mixture),
absorbed by a column density of $6\times 10^{21}$\,cm$^{-2}$ (for a
justification see below). The relationship between the volume density
of the hot gas and its measured emission measure and flux (see the
surface brightness measurements Sect.\,\ref{sec:diffuse_reduc}) was
obtained by a simulation of the adopted plasma adjusted to the
observational parameters and calibration. The volume density of the
outermost shell is used as the starting value for an iteration towards
the center of NGC\,3077; the $l(k)_{i}$ were re--calculated for each
annulus $k$ and shell $i$. Fig.\,\ref{fig:density} shows the resulting
three dimensional radial volume density profile of the hot gas in
NGC\,3077. The inset of this figure is an example for the method
described using six shells.

\placefigure{fig:density}

We fitted two models to the density profile (see also
Sect.\,\ref{sec:bubbles}): an exponential ($n(r)=a\,e^{-r/h}$) and a
power law model ($n(r)=(r/r_{c})^{-\beta}$). The best fitting
exponential has a scalelength of $h=271\pm 22$\,pc
($\chi^{2}_{red}=5\times10^{-4}$), whereas the best power law has an
index of $\beta=0.57\pm 0.06$ ($\chi^{2}_{red}=17\times10^{-4}$).

\subsection{Evidence for Expanding H$\alpha$ Shells Filled with Hot Gas} 
\label{sec:bubbles}

A comparison of the raw photon maps with an H$\alpha$ image shows that
expanding H$\alpha$ shells surround certain areas where diffuse X--ray
photons are detected. This is also stressed by the adaptively smoothed
representation shown in Fig.\,\ref{fig:color}b. The situation is
similar to what is expected from warm photoionized shells confining
hot bubbles \citep[][]{wea77}. The morphological relationship of the
images in the two different wavelengths is not due to spurious
artifacts of the adaptive smoothing process. A convolution of the raw
X--ray image with a fixed Gaussian kernel of $\sim 10\arcsec$ is very
similar to the results shown in Figs.\,\ref{fig:bands} and
\ref{fig:color}.

For a more quantitative analysis we constructed an intensity plot of
azimuthally arranged segments based on the \emph{unsmoothed} X--ray
data and an H$\alpha$ map. In order to avoid confusion with bright
\ion{H}{2} regions, the inner 20\arcsec\ of NGC\,3077 were excluded
from the analysis. Some of the segments cover regions where the
H$\alpha$ morphology reveals expanding shells and some are in between.
Fig.\,\ref{fig:cake} supports the existence of H$\alpha$ bounded hot
bubbles: e.g., segments 7, 8, and 9 are more X--ray luminous than the
surrounding segments 6 and 10/11 where the H$\alpha$ emission peaks.
In contrast, the latter two segments cover regions where the rim of
one particular expanding superbubble (R2, see Fig.\,\ref{fig:regions})
is present. The same behavior is found for segment 5 where the X--ray
intensity increases while the H$\alpha$ flux goes down. Segments 12 to
3 to the south east of NGC\,3077 are strongly affected by absorption
(see Sect.\,\ref{sec:various_x}) which makes a similar comparison more
difficult. The confinement of the hot gas is corroborated by the
morphology of an \ion{H}{1} feature to the west of NGC\,3077 which
resembles the rim of a supergiant shell and which appears to be a
barrier for the hot, coronal gas (see Fig.\,\ref{fig:color}c).

\placefigure{fig:cake}

We can also test this hypothesis by examining the density profile of
the hot gas: For a spherically--symmetric, outflowing wind,
theoretical models predict $n\propto r^{-\beta}$ with $\beta=2$ and
$T\propto r^{-4/3}$ \citep[$n$: volume density, $r$: distance along
outflow, $T$: temperature;][]{che85}. The same dependencies would be
observed for a conical outflow with a constant opening angle (S97).
A cylindrical geometry, however, results in a constant density all
along the outflow ($\beta=0$). As we show in Fig.\,\ref{fig:density},
a power law with an index of $\beta=2$ does not fit the volume density
profile of the hot gas detected in NGC\,3077. As dwarf galaxies have
low gravitational potentials, one naturally would expect that these
systems are to first order spherically--symmetric, but even the best
fitting power law ($\beta=0.57$, Sect.\,\ref{sec:diffuse_reduc}) is a
poorer fit compared to an exponential decline. In addition, we fitted
MeKaL and RS thermal plasma models to a circular region 10\arcsec\ in
radius centered on NGC\,3077 and to three surrounding consecutive
annuli with the same widths. The resulting temperatures varied by
$\sim 30$\% around $2.1\times 10^{6}$\,K but no clear radial
temperature gradient within the hot outflow is measured. This implies
that it is unlikely that NGC\,3077 currently forms a powerful hot
galactic outflowing wind similar to what is witnessed, e.g., in M\,82
\citep[well fit by a power law with an index of $\beta=0.9$ and with
varying opening angle][]{hec90,bre95} or NGC\,253 ($\beta=1.3$). We
conclude that the hot gas is indeed confined to giant H$\alpha$
shells.

\placefigure{fig:regions}

Based on this analysis we consider it justified to divide the diffuse
X--ray emission of NGC\,3077 in regions (R2--R7) coinciding with
expanding supergiant shells (see Fig.\,\ref{fig:regions}). One region
(R1) is defined to contain all of the X--ray emission within
NGC\,3077. We extracted ACIS--S3 spectra for all regions. Given their
positions on the CCD, all spectra were calibrated using the
appropriate RMFs and ARFs. Additionally, we defined background regions
with the same sizes and on similar CCD rows. To check how
significantly the calibration varies over the size of the aperture, we
also constructed intensity--weighted response files. The differences
to the spectra where the calibration files were not weighted were
found to be negligible. The total X--ray emission of NGC\,3077 (R1)
has enough counts to allow $\chi^{2}$ statistics to work properly. We
were therefore able to simply subtract the corresponding background
counts from the source data. In contrast to R1, the number of counts
per spectral bin from regions R2 to R7 is much lower and we applied
the Cash maximum likelihood algorithm accounting for Poisson
statistics. As a consequence, we were not able to subtract the
background from the data and we performed simultaneous fits of the
background (power law $\equiv$ powlaw1d) and of the source plus
background emissions ([photoelectric absorption $\times$ RS model] +
power law). In addition, we fitted MeKaL thermal plasma models to the
spectra \addtocounter{footnote}{-1} (\citealt{mew85,kaa92} and
updates\footnotemark\ with Fe--L calculations from \citealt{lie95}):
([photoelectric absorption $\times$ MeKaL model] + power law). The
errors of the fits were again obtained by simulations as is described
for the point sources in Sect.\,\ref{sec:3077point}. The differences
of the RS and the MeKaL models are considered to provide a measure for
systematic errors. The optimization method was the same as for the
point sources (Monte--Powell) and again we adopted solar metallicity
and element mixture. The resulting best fit parameters are tabulated
in Table\,\ref{tab:regions} ($N_{H}$: absorbing column density, $T$:
temperature, $F_{X}^{abs}$: absorbed flux, $F_{X}$: unabsorbed flux,
$L_X$: unabsorbed X--ray luminosity; errors are given at a 68\%
confidence level). The derived values for $N_{H}$ are rather high
while temperatures are low. Unfortunately, this has the effect that
the conversion from absorbed to unabsorbed fluxes is achieved by
rather large correction factors ($\sim 30-100$). The derived
unabsorbed X--ray fluxes and luminosities are therefore to be taken
with caution. Note that plotting the hardness ratios of the different
regions in a hardness ratio diagram (the shaded region in
Fig.\,\ref{fig:hardness}) reveals that the diffuse emission is very
unlikely to be due to a faint population of underlying point sources
with power law spectra. Thermal plasma models, however, are in good
agreement with the observed hardness ratios.

\placetable{tab:regions}

It is difficult to decide, which plasma model is a better fit to the
spectra of the individual shells. Even for the best fits to R1 (the
entire galaxy) goodness--of--fit indicators yield very similar values
for both the RS and the MeKaL model. In Fig.\,\ref{fig:spectra}c and d
we show this spectrum with the best fits overlaid. Additionally, we
calculated confidence regions for both models
(Fig.\,\ref{fig:spectra}b): the models agree well within a temperature
range of $T=0.15-0.25$\,keV (equal to $1.7-2.9\times 10^{6}$\,K) and
an absorbing column density range of $N_{H}=5-8\times
10^{21}$\,cm$^{-2}$ (boundaries of the 68.3\% confidence levels).
Similar plots are shown in the lower panels of
Fig.\,\ref{fig:contourplots} for the shells R2 to R7.

The best fitting absorbing column densities are quite high for the
entire galaxy compared to its measured \ion{H}{1} column density
(mean \ion{H}{1} column density: $\sim 2\times 10^{21}$\,cm$^{-2}$,
peak: $\sim 4\times 10^{21}$\,cm$^{-2}$, W02). However, molecular gas
is present in NGC\,3077 and can be detected by CO transitions at the
center. In addition, the confidence regions of the best fit to R1
(Fig.\,\ref{fig:spectra}b) shows that at least the MeKaL model
constrains the lower limit for absorption very poorly and absorbing
columns as low as say $10^{21}$\,cm$^{-2}$ are well within the
1\,$\sigma$ uncertainty. The RS model, however, sets a lower
1\,$\sigma$ limit of $\sim 5\times 10^{21}$\,cm$^{-2}$ which is
difficult to reconcile with \ion{H}{1} and CO data.

\placefigure{fig:spectra}
\placefigure{fig:contourplots}
\placefigure{fig:hardness}

From the normalization of the spectra (see footnote a in
Table\,\ref{tab:pointsources}), we were able to derive the densities
and the emission measures of each shell, assuming that the source
geometry is spherical. It is also assumed, that the electron and the
proton densities are equal, averaged over the entire volume
($<n_{e}>\approx<n_{p}>$). Note that the measured densities $n_{rms}$
must be converted to $n_{e}$ by using the appropriate filling factor
$f_{v}$ which is defined as $n_{rms}^{2}=f_{v} n_{e}^{2}$. The
pressures $P/k$ of the shells can then be derived via
$P/k=2\,n_{e}\,T$ ($k$: Boltzman's constant, $T$: temperature). In
addition, the hot gas mass $M_{hot}=n_{e} m_{p} V$ ($m_{p}$: proton
mass, $V$: shell volume), the thermal energy $E_{th}=3 n_{e} V kT$,
the cooling timescale $t_{cool}=E_{th}/L_X$, the mass deposition rate
$\dot{M}_{cool}=M_{hot}/t_{cool}$, and the mean particle velocity
$<v_{hot}> = \sqrt{2 E_{th}/M_{hot}}$ are listed in
Table\,\ref{tab:gas} which also shows their dependence on $f_{v}$ and
on the metallicity of the hot gas ($\xi$, in solar metallicity units).
The equivalent diameters $d_{eq}$ were derived as being the diameters
of circles which have the same areas as the regions under
consideration ($d_{eq}=\sqrt{4\,Area/\pi}$). The mean lines of sight
are obtained by $2/3\times d_{eq}$ (cylindrical projection). Errors
imposed on the physical gas parameters by the source geometry are
estimated to be of order 30\%.

\placetable{tab:gas}

For the individual shells we derive temperatures of $\sim 1.3$ to
$4.9\times 10^{6}$\,K, hot gas masses of order $\sim
10^{5}\,\xi^{-0.5}\,f_{v}^{0.5}$\,M$_{\sun}$, and particle volume
densities of $\sim 0.05-0.9\,\xi^{-0.5}\,f_{v}^{-0.5}$\,cm$^{-3}$. The
pressures of the hot gas derived for the individual shells in
NGC\,3077 are $P/k\approx 2-30\times
10^{5}\,\xi^{-0.5}\,f_{v}^{-0.5}$\,K\,cm$^{-3}$. This is in contrast
to typical pressures of the ISM in the Milky Way which are of order a
few thousand \citep[e.g.,][]{wol95}. It is this overpressure of the
hot gas which drives the expansion of the superbubbles.

Taking the hot gas masses $M_{hot}$ and the age of the shells
$t_{shell}$ (Sect.\,\ref{sec:theoX} and Table\,\ref{tab:gas}), we can
estimate the rate of hot gas deposed to the halo. We derive
$M_{hot}/t_{shell}\approx 0.1-0.5
\,\xi^{-0.5}\,f_{v}^{0.5}$\,M$_{\sun}$\,yr$^{-1}$ which is similar to
the warm, ionized gas ($M_{warm}/t_{shell}=0.6\times
(f_{v,warm}/0.1)^{0.5}$\,M$_{\sun}$\,yr$^{-1}$, M98) and a few times
greater than the current star formation rate ($\sim
0.06$\,M$_{\sun}$\,yr$^{-1}$, W02; value adapted to a distance of
NGC\,3077 of 3.6\,Mpc).

The value of the filling factor $f_{v}$ is uncertain. Hydrodynamic
simulations and Chandra observations of superwind galaxies
\citep{str00,str02} show that $f_{v}$ is about 0.1--0.3. The hot gas
in NGC\,3077, however, is still confined to giant shells and the
X--ray emission emerging from these regions does not appear to be
brighter in the vicinity of the H$\alpha$ emitting rims as compared to
the interior of the shells (see above). A more quantitative criterion
is that the time $t_{c}$ needed to cross a superbubble with the sound
speed $c$ of the plasma is less than the age of the shell. The sound
speed is given by $c=\sqrt{2\gamma k T/m_{p}}$ where $\gamma=5/3$
(S97). Taking a mean temperature of $2\times 10^{6}$\,K and a
diameter of the bubble of $300$\,pc (see Tables\,\ref{tab:regions} and
\ref{tab:gas}) we derive $c\approx 240$\,km~s$^{-1}$ and $t_{c}\approx
1$\,Myr. $t_{c}$ is a factor 2--10 less than the ages of the bubbles
and the hot plasma had enough time to thermalize within the cavities.
A volume factor close to unity therefore appears to be more satisfying
than the lower ones detected in outflows.

There are some differences in the properties of the individual shells.
R7 and R4 were fitted to have high absorbing column densities and low
temperatures relative to the others. The combination of these
parameters lead to relatively high correction factors for absorption
and therefore high unabsorbed fluxes and luminosities. The opposite is
the case for R3 and R6, the best fits suggest lower absorbing column
densities and higher temperatures. The highest number of photon counts
is observed in R2 which corresponds to the highest absorbed flux for
all shells. However, after correcting for absorption, the unabsorbed
X--ray luminosity is in the mid range of the shell luminosities. In
addition, R7 and R4 have the highest normalizations which results in
relatively high volume densities, masses, thermal energies and
pressures of the hot gas. Again, R6, R3 and also R5 show relatively
low values for these parameters. The reason for this might be that R7
and R4 are located very close to the stellar disk of NGC\,3077 (see
Fig.\,\ref{fig:color}b) which is centered on R3 with a position angle
of $\sim 45$\degr\ \citep[taken from the LEDA catalog][]{prug98}. In
this case, R3 itself, which seems to be absorbed by a layer of rather
low column density, must be located closer to the observer as compared
to R4 and R7, in particular in front of the extended \ion{H}{1}
feature in the south--east and the molecular gas in the central region
of NGC\,3077 (see Fig.\,\ref{fig:color}c). Alternatively, NGC\,3077
maybe inclined in such a way that the far side of the disk of
NGC\,3077 points towards the south--east. However, the distribution
of \ion{H}{1} within NGC\,3077 is rather chaotic and is not in the
form of a regular disk along with its optical counterpart. Significant
contamination with non--thermal X--ray photons is not supported by the
spectral data of R3 (and all the other shells); above an energy of
$\sim 1.5$\,keV no emission exceeding the background can be detected
(see also the shaded region in the hardness ratio diagram
Fig.\,\ref{fig:hardness}).

\subsection{The Total Spectrum}
\label{sec:total}

We extracted a total X--ray spectrum of NGC\,3077 including both point
sources and diffuse emission (Fig.\,\ref{fig:tot_spectrum}). In order
to verify the results obtained for the individual spectral fits, we
added the following best fit model components and compared the results
to our measurement. Point sources: RS plasmas for S1, S5, and S6,
power law for S2 and S3, black body for S4 (see
Sect.\,\ref{sec:discuss_pointsources} and
Table\,\ref{tab:pointsources}); Diffuse Emission: RS/MeKaL plasma for
R1 (see Tab.\,\ref{tab:regions}). For all models, we adopted the
absorbing column densities as derived by the corresponding fit. For
energies $\lesssim 0.8$\,keV and $\gtrsim 1.2$\,keV the total model
provides a good fit to the data. In the $0.8$\,keV$\lesssim E \lesssim
1.2$\,keV energy range, the total model shows a count rate which is
$\sim 5\times 10^{-3}$\,cts~s$^{-1}$~keV$^{-1}$ too high. This slight
offset is most likely due to the fits to the point sources, where we
modeled the background by a power law. In Fig.\,\ref{fig:tot_spectrum}
we show the added best fitting RS/MeKaL plasma models for the regions
R2 to R7 as well. The derived combined flux of the regions R2 to R7 is
a factor of $\sim 1.6$ lower than the diffuse X--ray flux of the
entire galaxy (R1) in the case of both, the RS and the MeKaL models.
The total unabsorbed luminosity of NGC\,3077 is $2-5\times
10^{39}$\,erg~s$^{-1}$, depending on the plasma model used.

\placefigure{fig:tot_spectrum}

\subsection{Is there an X--ray Population in the Tidal Tail close to NGC\,3077?}
\label{sec:field}

NGC\,3077 finds itself in interaction with M\,81 and M\,82 (the M\,81
triplet of galaxies). Large \ion{H}{1} tidal tails cover the region
around these galaxies \citep[see, e.g.,][]{yun94}. To the east of
NGC\,3077, a massive tidal \ion{H}{1} complex is present and molecular
gas and star formation as traced by CO and \ion{H}{2} regions have
been found in this tidal feature (e.g., see \citealt{wal99}, W02,
\citealt{kar85}). The roll angle and the aimpoint of the Chandra
observations presented here were carefully chosen in a way such that
in addition to NGC\,3077 the bulk of the \ion{H}{1} emission and most
of the extragalactic \ion{H}{2} regions are within the ACIS--S3 field
of view (see Fig.\,\ref{fig:field}).

In addition to the point sources associated with NGC\,3077, we detect
38 point sources in the ACIS--S3 field of view, 24 of which were
detected by both the {\it wavdetect} and the {\it celldetect}
algorithm (see Sect.\,\ref{sec:detect}). We compared the source list
(Table\,\ref{tab:field}) to the SIMBAD database and found that three
sources correspond to known stars: two form the bright double star
HD\,8667 to the north and one source is coincident with a star
detected by the Hubble Space Telescope pointing at the halo of
NGC\,3077 \citep{sak01}. The counts for each X--ray point source and
band are listed in Table\,\ref{tab:field}.

Fig.\,\ref{fig:hardness} shows a hardness ratio plot of all point
sources in the field of view which were detected by both source
detection algorithms. Most of the sources are consistent with modeled
power law spectra. Clearly offset are the three detected stars as well
as S4, S5, and S6 within NGC\,3077, which are substantially softer
(see Sect.\,\ref{sec:discuss_sn} and
\ref{sec:discuss_sss}). This implies that we find no evidence for 
a bright population of soft sources in the tidal tail (e.g., SNRs).

To check whether the detected X--ray sources belong to the M\,81
triplet or if they are background objects we compared the data to the
results from the Chandra Deep Field South project \citep[CDFS;][in the
following G01]{gia01}. To do so, we re--reduced our Chandra data in
the same manner as G01. Within two new energy bands (G01 soft:
0.5\,keV$\leqslant E \leqslant$2.0\,keV, G01 hard: 2.0\,keV$\leqslant
E \leqslant$7.0\,keV), the data were binned to a pixel size of
1\arcsec\ and were corrected for vignetting effects. Fluxes were
calculated regarding the Galactic foreground column density along the
line of sight to NGC\,3077. For a full description of the data
reduction process see G01.

Excluding the identified stars and the sources within NGC\,3077, we
constructed $\log N(>S) - \log S$ plots ($N$: cumulative number of
sources brighter than $S$ per square degree, $S$: flux in
erg~s$^{-1}$\,cm$^{-2}$) for both of the G01 bands. In
Fig.\,\ref{fig:logn} we compare our results to those obtained by G01
for the CDFS. All data are well described by the G01 relations. We do
not observe a significant overdensity of X--ray sources within the
tidal tail around NGC\,3077 down to our detection limits. This is in
agreement with scaling the number of the point sources within
NGC\,3077 by the ratio of blue and H$\alpha$ luminosities of NGC\,3077
and the ``Garland'' region \citep[see][]{kar85} which occupies a
considerable fraction of the tidal tail
($L_{B}^{Garland}/L_{B}^{NGC3077}=5.2\times
10^{6}$\,L$_{B\sun}/1.1\times 10^{9}$\,L$_{B\sun}\approx 1/200$;
$F_{H\alpha}^{Garland}/F_{H\alpha}^{NGC3077}=1.88\times
10^{-13}$\,erg~s$^{-1}~$cm$^{-2}/3.75\times
10^{-12}$\,erg~s$^{-1}~$cm$^{-2}\approx 1/20$; \citealt{sha91},
\citealt{dvo91}, \citealt{wal03}). Assuming that these ratios are a
measure for the ratio of the number of X--ray point sources, we expect
to detect no X--ray point source in the Garland region. This estimate
is in agreement with our analysis and in contrast to the results of
\citet{bi94}. They find an overpopulation of X--ray sources in the
tidal tail of a factor of two, based on their 7\,ks ROSAT data of a
field centered on NGC\,3077 and 10\arcmin\ in size (roughly
corresponding to the ACIS--S3 field of view).

\placefigure{fig:logn}

\section{Discussion}
\label{sec:discuss}

\subsection{The Nature of the Discrete X--ray Sources in NGC\,3077}
\label{sec:discuss_pointsources}

\subsubsection{Supernova Remnants}

\label{sec:discuss_sn}

The spectrum of S1 (Fig.\,\ref{fig:p1}) shows identifiable line
complexes. S1 is spatially coincident with radio continuum emission, a
CO peak and a strong absorption feature in the optical. The Pa$\alpha$
\citep{boe99} and the radio continuum morphology are 
similar, which suggests that the radiation arises from the same region
and is due to shocked gas. This would agree with the fact that we
derive a radio continuum spectral index of $\alpha=0.48$ \citep[def.:
$S \propto \nu^{-\alpha}$, data from][]{nik95,con87} which is
consistent with spectral indices found by \citet{hen01} for a sample
of SNRs in the Large Magellanic Cloud.

At the distance of NGC\,3077, the slightly extended source S5 has a
size of $19$\,pc$\times $23\,pc. The hardness ratio plot
(Fig.\,\ref{fig:hardness}) reveals, that S5 and S6 are rather soft
sources compared to background objects and to power law models. From
the fitting process we cannot reliably decide which of the three
models (RS, PL, BB) provides the best fit to the data of S1, S5, and
S6. However, the photon indices of the PL model are too high to be
explained by any known celestial source. In addition, the alternative
BB models yield temperatures which are in excess of at least a factor
of two to models of white dwarf or isolated neutron stars
\citep[see][]{kah97,kah99,hab99,bur01,zam01}. For these reasons, we conclude 
that S1, S5, and S6 may be (young) SNRs and emit X--rays from a
thermal plasma. They are close to the actively star forming regions in
NGC\,3077 and to the molecular complexes (see Fig.\,\ref{fig:color}).
From the best fitting RS and MeKaL plasma models we derive the
parameters given in Tables\,\ref{tab:pointsources} and
\ref{tab:sn_sources}.

\placetable{tab:sn_sources}

\subsubsection{Accreting Objects}

\label{sec:discuss_plaw}

S2 and S3 exhibit a somewhat similar spectrum with no clear energy
peak. The best fitting RS and BB spectra result in high temperatures.
PL fits are supported by the hardness ratio plot
(Fig.\,\ref{fig:hardness}) where the hardness ratios of S2 and S3 are
similar to those derived for background objects and are consistent
with simulated PL models. The fitted absorbing column densities differ
by a factor of $\sim 3$; it is likely that they are unrelated objects,
in spite of their small angular separation ($1\arcsec$). Due to the
low S/N, we can only speculate that these sources may be accreting
objects. Judging from their luminosities, the nature of these objects
can be both, low or high mass X--ray binaries. Both sources might be
background AGNs, too. Especially S2 might be such an object, due to
the steepness of its spectral index ($\gamma=1.65$; cf. sources in the
CDFS: $\gamma=1.70\pm0.12$, see G01) and its high absorbing column
density which might place S2 behind NGC\,3077. Statistically, given
the density of background sources in a typical field (e.g., the CDFS),
a background object coinciding with the diffuse emission of NGC\,3077
is possible.

\subsubsection{The Supersoft Source}

\label{sec:discuss_sss}

The point source with the softest spectrum is S4. It is located on the
rim of an H$\alpha$ feature. Its lack of line features and emission
harder than 0.8\,keV suggests that the spectrum is best described by
a black body model ($T=9.4\times 10^5$\,K equivalent to 81\,eV). A
power law index of 5 would be too steep for any Galactic or
extragalactic source and the best RS plasma fit features the Fe--L
line complex near 0.9\,keV which is not observed. The temperature and
the X--ray luminosity of $9.2\times 10^{36}$\,erg~s$^{-1}$ of the BB
spectrum puts this object within the range found for other supersoft
sources \citep{kah97,swa02}.

It is possible to derive the radius of an object from the amplitude of
the BB spectrum (see footnote b in Table\,\ref{tab:pointsources}). For
the distance of NGC\,3077 we find a radius of $R\sim1600$\,km. This is
much lower than the radius of a typical supersoft source, a hydrogen
burning white dwarf \citep[WD; $R\sim
6000-9000$\,km;][]{kah97}. However, since the hydrogen burning phase
is confined to a small region on the surface of the WD, the derived
value might be in accordance with such an object.

An alternative interpretation would be that S4 is an isolated neutron
star. The radius of a typical neutron star is about 20\,km. If we
adopt this typical radius, then S4 would have a distance of only
45\,kpc, which puts it in the halo of the Milky Way. Located at this
distance, its luminosity would drop to $L_X = 1.4 \times
10^{33}$\,erg~s$^{-1}$ which is rather high for an isolated neutron
star but not impossible \citep[see, e.g.,][]{zam01}. The best fit for
the column density is similar to the Galactic foreground hydrogen
content and is even consistent with a value of zero within the errors
(Fig.\,\ref{fig:p23456}).

One can estimate the expected number of supersoft sources by a
comparison of NGC\,3077 with M\,81, its neighbor galaxy. M\,81 was
observed with Chandra's ACIS--S3 CCD for about 50\,ks. This
observational setup and integration time is comparable to the X--ray
observations discussed here. Based on that data, \citet{swa02} find 9
supersoft sources in M\,81. The absolute blue magnitudes \citep[M\,81:
$-19.8\pm0.14$\,mag, NGC\,3077: $-17.1\pm0.14$\,mag;][]{dvo91} show
that M\,81 is about a factor of 10 more luminous than NGC\,3077.
Scaling the number of supersoft sources by this factor, one would
expect one supersoft source in NGC\,3077 which is what is observed.

\subsection{ The Diffuse X--ray Emission of NGC\,3077}
\label{sec:dis_diffuse}

\subsubsection{Pressure--Driven Bubbles}
\label{sec:theoX}

Theoretical models of expanding supershells are generalized models of
winds from massive stars streaming into the ambient ISM \citep[see,
e.g.,][]{cas75,wea77,mac88}. In the following, we will use the set of
equations derived by \citet{chu90} who used the EINSTEIN band (0.2 --
4.0\,keV) over which to calculate the X--ray emissivity of hot
gas. For temperatures of about $2\times 10^6$\,K, no substantial
emission above 4.0\,keV is expected, so their results should be
applicable to the Chandra data discussed here. The theoretical X--ray
luminosity of a pressure--driven superbubble is given by

\begin{equation}
L_X=3.29\times 10^{34} {\rm erg~s}^{-1}\,\xi \,I(\tau)\,
L_{mech, 37}^{33/35}\,\, n_{amb}^{17/35}\,\, t_{shell, 6}^{19/35}
\label{eq:lx_theory}
\end{equation} 

\noindent where $\xi$ is the metallicity of the hot gas with respect to solar, 
$L_{mech, 37}$ the mechanical luminosity in units of
$10^{37}$\,erg~s$^{-1}$, $n_{amb}$ the number density of the ambient
material in cm$^{-3}$, and $t_{shell, 6}$ the lifetime of the starburst in Myr. The
integral $I(\tau)=(125/33) - 5\,\tau^{1/2} + (5/3)\, \tau^3 - (5/11)\,
\tau^{11/3} $ only depends on the dimensionless temperature
$\tau=T_{min}/T_{center}$ given a cutoff temperature $T_{min}$ and the
temperature at the center of the shell $T_{center}$. Expressed in
observables:

\begin{equation}
\tau = 0.16\, L_{mech,37}^{-8/35}\,\, n_{amb}^{-2/35}\,\, t_{shell,6}^{6/35}.
\end{equation}

\noindent The ambient density $n_{amb}$ is related to the electron 
density in the interior of the shell $n_{e}$ by

\begin{equation}
n_{amb} \approx \frac{n_{e} k T}{m_{p} v_{exp}^{2}}
\label{eq:amb_dens}
\end{equation}

\noindent where $k$ is Boltzman's constant, $T$ the temperature, $m_p$ the
proton mass, and $v_{exp}$ the expansion velocity of the shell.

The density--dependent mechanical luminosities $L_{mech}/n_{amb}$,
the ages $t_{shell}$, and the velocities $v_{exp}^{proj}$ of the
shells were previously derived by M98 using long--slit echelle spectra
(see Table\,\ref{tab:gas}). Adopting solar metallicity and the best
fit values from Table\,\ref{tab:regions}, $I(\tau)$ is about 2.8 and
the resulting predicted densities of the ambient material, the
mechanical luminosities $L_{mech}$, and the predicted X--ray
luminosities are listed in Table\,\ref{tab:gas}. Note, that shells A
and J in M98 have a somewhat larger size compared to R3 and R7.
However, this should not have any significant impact on the predicted
numbers.

A comparison of the theoretical values with the ones directly derived
from the Chandra data (Tables\,\ref{tab:regions} and \ref{tab:gas})
reveals that the observed luminosities are in reasonable agreement
with the basic theory developed by \citet{wea77}. In their model,
``boiled--off'' material from the rim of the superbubble is the only
mechanism for mass--loading. Therefore, the hot gas which fills the
cavities is expected to originate only from the evaporation process of
the rim as well as from the stellar ejecta themselves. We estimate the
masses of both processes. \citet{wea77} derive the evaporated mass
rate $dM_{b}/dt$ to

\begin{equation}
\frac{dM_{b}}{dt}\,=\,C\, <T>^{5/2}\, \frac{R^{2}}{R-r}
\label{eq:weaver_boil}
\end{equation}

where $C=4.13\times 10^{-14}$ (cgs units) is a material--dependent
constant, $<T>$ is the mean temperature of the hot gas, $R$ the radius
of the superbubble, and $r$ the radius of a sphere where the stellar
ejecta has not thermalized yet. Given the large sizes, ages, and the
plasma sound speed of the superbubbles in NGC\,3077 (see
Sect.\,\ref{sec:bubbles} and Table\,\ref{tab:gas}) we expect that all
the material is already thermalized, i.e., $r=0$. By using the derived
values given in Table\,\ref{tab:gas} and integrating over the age of
the individual bubbles we estimate the amount of the evaporated mass
to $\sim 20-60$\% for R3, R5, and R6 and $\sim 1-10$\% for R2 and R7
of the hot gas masses. The uncertainties are given by the different
plasma models.

In addition, we used the STARBURST\,99 models provided by
\citet{lei99} to estimate the injected mass from the combined effect
of stellar winds and SNe. To do so, we estimate the star formation
rate (SFR) within each superbubble by scaling the measured mechanical
luminosity (Tab.\,\ref{tab:gas}) by the asymptotic values of the
STARBURST\,99 models with continuous star formation (Salpeter IMF,
upper mass cutoff: 100\,M$_{\sun}$, lower mass cutoff: 1\,M$_{\sun}$,
solar metallicity). The SFRs over the lifetimes of the superbubbles
are derived to $0.02-0.06$\,M$_{\sun}$~yr$^{-1}$ for each superbubble
except for R7 ($\sim 0.4$\,M$_{\sun}$~yr$^{-1}$). These values are
subsequently used to scale the mass loss predicted by the theoretical
models integrated over the ages of the individual superbubbles. As a
result we estimate that about $30-50$\% of the hot gas within R3, R5,
R6, and R7, and about $1-10$\% within R2 may originate directly from
stellar ejecta.

A mass balance shows that additional mass--loading besides evaporation
of the rims and stellar ejecta, e.g., by evaporation of halo or disk
clouds, is definitely needed for the largest superbubble R2 in order
to account for $\gtrsim 80$\% of its hot gas mass. R7 needs some $\sim
40$\% of additional mass--loading and the other superbubbles R3, R5,
and R6 can be explained entirely without this mechanism.

Adding up the estimated SFRs of the superbubbles obtained by the
STARBURST\,99 models yields $\sim 0.6$\,M$_{\sun}$~yr$^{-1}$ which is
about an order of magnitude higher than the current SFR
(0.06\,M$_{\sun}$\,yr$^{-1}$, W02, based on H$\alpha$ and Pa$\alpha$
flux measurements). The hot and warm gas deposition rates lead to
similar discrepancies with the current SFR (Sect.\,\ref{sec:bubbles}).
Note that the 1.4\,GHz radio continuum flux of W02's VLA data
($\sim42$\,mJy) results in a similar current SFR of
$\sim0.08$\,M$_{\sun}$\,yr$^{-1}$ \citep[conversion factor taken
from][]{haa00}. Also, the far--infrared luminosity of NGC\,3077
\citep[$3.2\times 10^{8}$\,L$_{\sun}$,][adapted to a distance of
3.6\,Mpc]{yun99} can be converted to a SFR of
$\sim0.06$\,M$_{\sun}$\,yr$^{-1}$ \citep[conversion factor taken
from][]{ken98}. The differences of the STARBURST\,99 and current SFR
estimates might indicate that the star forming activity of NGC\,3077
has decreased with time over the last few Myr (for the ages of the
shells, see Table\,\ref{tab:gas}).

\subsubsection{Impact of Superbubbles on the Intragroup Medium}

The superbubble shells will accelerate at radii where the ambient gas
density drops rapidly. As hydrodynamic instabilities break up the
shells, the hot gas will rush outwards in a galactic wind. An obvious
question is whether the outflowing material will return to NGC\,3077 or
not. The answer depends on several factors. A necessary condition is
that the wind has enough energy to escape from the gravitational
potential. The wind can still stall, however, if the hot wind cools
and/or does a great deal of work pushing an extended gaseous, galactic
halo aside. The cooling times of the shells (2--20\,Myr, see
Table\,\ref{tab:gas}) exceed the flow time (1--10\,Myr) so radiative
losses do not direct the dynamical evolution at this stage. Hence, we
discuss only the impact of the gravitational potential and gaseous
halo.

The tidal interaction with M\,81 and M\,82 \citep[e.g.,][]{yun99}
raises questions about the reality of mass models for NGC\,3077. M98
assume a \ion{H}{1} rotation speed of 40\,km~s$^{-1}$ across the
system. However, higher angular resolution \ion{H}{1} data (W02)
subsequently showed the \ion{H}{1} velocity spread within the optical
body of NGC\,3077 to be 90\,km~s$^{-1}$ (1$\sigma$ velocity dispersion
of $\sim 25$\,km\,s$^{-1}$). W02 detect $6.1\times 10^{8}$\,M$_{\sun}$
of \ion{H}{1} in the tidal tail (including NGC\,3077) but only
$1.3\times 10^{8}$\,M$_{\sun}$ of \ion{H}{1} in NGC\,3077 (values
adjusted to a distance of 3.6\,Mpc), and no rotation was detected in
the latter. The stellar mass of NGC\,3077 is estimated to be
$1.4\times 10^{9}$\,M$_{\sun}$ based on the blue luminosity and
$M/L_{B}=1$ \citep{dvo91}. Hence, the total baryonic mass is probably
$1.5-2\times 10^{9}$\,M$_{\sun}$.

Within a simple model, we scale the dark mass $M_{DM}$ of NGC\,3077 by
its visible mass $M_{vis}$ according to the empirically derived
relation \citep{per96,sil01}

\begin{equation}
M_{DM}=3.74\times 10^{8} \left(\frac{M_{vis}}{10^{7} M_{\sun}}\right)^{0.71}\,{\rm M}_{\sun} 
\end{equation}

and estimate $M_{DM}\approx 10^{10}$M$_{\sun}$ leading to
$M_{DM}/M_{vis}\sim 10$ \citep[in agreement with the estimate
of][based on $M/L\approx9$]{bro91}. The complicated history of
NGC\,3077, which might have had a much larger (or smaller) mass in the
past, however, amplifies the uncertainty in the ratio of dark to
baryonic matter.

Given an isothermal dark matter halo, the escape velocity is a
function of radius $r$ and rotation speed $v_{rot}$. It is described
by

\begin{equation}
v_{esc}= v_{rot}\,\sqrt{2\,\left(1+\ln \frac{r_{max}}{r}\right)}
\end{equation}

with a maximal halo radius $r_{max}$ of

\begin{equation}
r_{max}=0.016\,\left(\frac{M_{DM}}{{\rm M}_{\sun}}\right)^{1/3}\,\,h^{-2/3}\,{\rm kpc}
\end{equation}

\citep[$h$: dimensionless Hubble constant, 
adopted to $h=0.7$ ][]{mac99,fer00,sil01}. For NGC\,3077 we derive an
escape velocity of $\sim 110$\,km~s$^{-1}$ at the center and $\sim
80$\,km~s$^{-1}$ at 1\,kpc above the disk. Expressed as an escape
temperature, the bubble gas needs to exceed $T=2-9\times 10^{5}$\,K to
be assured of escape. This condition is clearly met by the hot
bubbles ($T\approx1-3\times 10^{6}$\,K). In contrary, the expansion
speed of the warm shells is about equal to the maximum escape velocity
predicted.

As a second condition, the expanding bubbles must provide the work to
push the halo gas aside. The velocity of a shell $v_{s}$ is a function
of the pressure of the hot gas $P_{hot}$ and the density of the
ambient medium $\rho_{amb}$: $v_{s}=(P_{hot}/\rho_{amb})^{0.5}$
\citep{koo90,sil01}. For decreasing density, e.g., a stratified
atmosphere, the velocity of a shell increases as long as the driving
pressure is constant. For an ultimate loss of material the density has
to decrease sharply before the hot gas in the interior of the shell
cools down. \citet{koo92} show that for an exponential density profile
the shell has to reach three times the exponential scalelength before
it re--accelerates, fragments, and ultimately blows the currently
stored hot gas out of the galaxy's gravitational potential. Assuming a
Gaussian stratification, this effect will start at about one Gaussian
scalelength.

The geometry of the \ion{H}{1} data is more complicated, but we argue
that escape is clearly easiest where the gradient in halo gas density
is steepest, which is roughly along the north--south axis (see
Fig.\,\ref{fig:field}). This is also the direction where the largest
elongation of the diffuse X--ray emission is detected
(Fig.\,\ref{fig:bands}). Towards the south, NGC\,3077 is connected to
a tidal tail which implicates that the \ion{H}{1} column density is
very extended ($\sim 5$\,kpc) at a constant $N_{HI}\approx 1.5\times
10^{21}$\,cm$^{-2}$ (Fig.\,\ref{fig:hi_profile}). Towards the north,
however, a continuous decline is observed. Hot gas streaming to the
south will most likely not be able to overcome the large ISM pressure
``barrier''; the cooling times of the hot gas within the shells are
much lower than the time needed to reach the outer boundary of the
southern tidal tail (see Table.\,\ref{tab:gas}). This is in agreement
with the morphology of the diffuse X--ray emission
(Fig.\,\ref{fig:bands}) which is considerably less extended towards
the south compared to the north.

\placefigure{fig:hi_profile}

We fitted a Gaussian and an exponential profile to the northern half
of the \ion{H}{1} column density profile of NGC\,3077 and derive the
respective scalelengths to $h_{gauss}\approx 0.5$\,kpc and
$h_{exp}\approx 0.6$\,kpc (for comparison: the scalelength of the
total diffuse X--ray emission is $\sim 0.2$\,kpc,
Sect.\,\ref{sec:diffuse_reduc}). Both profiles provide an acceptable
fit. Via $N_{HI}=\sqrt{2\pi}\,h_{gauss}\,n_{0}$ (Gaussian) and
$N_{HI}=h_{exp}\, n_{0}$ (exponential), we can derive the midplane
volume density $n_{0}$ of NGC\,3077. Assuming that the mean column
density is $N_{HI}\approx 2\times 10^{21}$, we consistently obtain
$n_{0}=0.5-1.0$\,cm$^{-3}$ for both models. This is in good agreement
with what we predict based on Eq.\,\ref{eq:amb_dens} for the ambient
densities of the shells (see Table\,\ref{tab:gas}). If the H$\alpha$
shells do not decelerate significantly (current maximum velocity:
$v\approx 100$\,km~s$^{-1}$), they will reach the \ion{H}{1}
scalelengths in $\sim 5$\,Myr just before substantial cooling takes
place (Table\,\ref{tab:gas}). In addition, current star formation
provides additional energy to the shells. It is likely that the shells
and the hot gas indeed reach the northern \ion{H}{1} scalelengths
while maintaining supersonic speed. Eventually, the shells will
fragment due to Rayleigh--Taylor instabilities and release their hot
gas to the intragroup medium of the M\,81 group of galaxies.

Recently, \citet{kar02} estimated the mass of the M\,81 group to
$3-16\times 10^{11}$\,M$_{\sun}$, its rotational velocity to $\sim
250$\,km~s$^{-1}$ and its size to $\sim 500$\,kpc. Assuming M\,81 to
be the center of the group (distance to NGC\,3077 $\sim 50$\,kpc), we
estimate the escape velocity of the group at the position of NGC\,3077
to $\sim 500-600$\,km~s$^{-1}$. This converts into an escape
temperature of $\sim 2-3\times 10^{7}$\,K, about an order of magnitude
higher than the current temperature of the hot gas: the hot gas of
NGC\,3077 is not able to escape from the M\,81 group of galaxies.

In summary, we conclude that some of the hot gas will likely enrich the
intragroup medium of the M\,81 group to the north but not to the south
of NGC\,3077. The IGM beyond the M\,81 group, however, will not be
affected by the developing superwind of NGC\,3077.

\subsection{Comparison to M\,81 and M\,82}
\label{sec:m82}

NGC\,3077 is interacting with its neighbors, the spiral galaxy M\,81
and the starburst galaxy M\,82. A compilation of some general and
X--ray properties of these three galaxies can be found in
Table\,\ref{tab:triplet} \citep[the ROSAT HRI X--ray data of M\,81 has
been taken from] [the ROSAT PSPC data of M\,82 from S97]{imm01}. In
contrast to NGC\,3077, both of its neighbors harbor a strong
point--like nuclear source, each of which is believed to be associated
with a black hole system \citep[e.g.,][]{iyo01,kaa01}. To make a
proper comparison of the X--ray properties, we exclude these sources
in the following discussion (resulting in columns 5 and 8 of
Table\,\ref{tab:triplet}). Furthermore, \citet{imm01} model the
diffuse emission of M\,81 by two thermal plasmas and a luminous power
law component. The latter is supposed to comprise faint point sources,
rather than truly diffuse gas, which is why we add this component to
the point source flux and extrapolate the luminosities accordingly
(column 6).

\placetable{tab:triplet}

The dominating X--ray emission emerging from M\,81 is in the form of
point sources (91\%). This is not observed in M\,82 and NGC\,3077
where most of the X--ray photons are radiated by hot outflowing gas
(95\% and 85\% respectively). The diffuse X--ray luminosity of
NGC\,3077 is more luminous than the one in M\,81 even in absolute
terms (Table\,\ref{tab:triplet}). As expected, the absolute X--ray
luminosity of the point sources is higher for more massive galaxies.

S97 discuss the physical properties of the galactic wind of M\,82.
They find a total hot gas mass of $M_{hot}=1.3 \times
10^{8}\,f_{v}^{0.5}$\,M$_\sun$, a thermal energy of $E_{th}=3.6 \times
10^{56}\,f_{v}^{0.5}$\,erg, and a mass deposition rate of
$\dot{M}_{cool}=0.2$\,M$_\sun$\,yr$^{-1}$ (NGC\,3077: $M_{hot}\approx
3 \times 10^{6}\,\xi^{-0.5}\,f_{v}^{0.5}$\,M$_\sun$, $E_{th}\approx 3
\times 10^{54}\,\xi^{-0.5}\,f_{v}^{0.5}$\,erg,
$\dot{M}_{cool}\approx0.1$\,M$_\sun$\,yr$^{-1}$, see
Table\,\ref{tab:gas}). The cooling times $t_{cool}$ of the coronal gas
in M\,82 are derived to be about $600\,f_{v}^{0.5}$\,Myr, and the
temperatures are some $5 \times 10^6$\,K (NGC\,3077: $t_{cool}\approx
40\,\xi^{-0.5}\,f_{v}^{0.5}$\,Myr, $T\approx 2\times 10^6$\,K). Point
source subtraction in the S97 data, however, is a difficult task,
given the low ROSAT resolution. Compared to M\,82, the starburst of
NGC\,3077 is weaker in absolute terms. The temperature of the hot gas
in NGC\,3077, however, is only a factor of two lower as compared to
that in M\,82 and very similar to the dominating hot plasma component
of M\,81 ($\sim 1.7\times 10^{6}$\,K). In addition, the pressure of
the hot gas in M\,82 is even a factor of $\sim 2$ lower as compared to
NGC\,3077. This is an effect of the $\sim 5$ times lower density
within the outflowing superwind in M\,82.

Normalized to the \ion{H}{1} contents or the blue luminosities of
NGC\,3077 and M\,82 (which are potential measures for the ``fuel'' of
the starbursts and the total masses of their host galaxies), the
strong galactic superwind in M\,82 still carries a higher mass and
thermal energy of the hot gas. The situation is different, however,
when normalizing these quantities to the current SFRs. The SFR
obtained from H$\alpha$ measurements, the far--infrared luminosity,
the thermal energy of the hot gas and its mass, all these parameters
are some two orders of magnitudes lower in NGC\,3077 as compared to
M\,82. This result is somewhat intriguing given the very different
physical states of the strong galactic superwind in M\,82 and the hot
gas confined in superbubbles in NGC\,3077. Note that the star formation
efficiency based on molecular gas ($SFR/M_{H_{2}}$) of NGC\,3077 and
M\,82 are approximately the same, too (W02).

\section{Summary and Conclusions}
\label{sec:summary}

In this study, we present deep (53\,ks) Chandra observations of the
dwarf starburst galaxy NGC\,3077 and its environment. Our results are
the following:

\begin{enumerate}

\item The main X--ray emission emerges from a diffuse soft 
    component 1\,kpc in size. In addition, 6 point sources are
    detected. The point sources contribute 8--21\% to the total
    X--ray luminosity, the diffuse emission 79--92\% (depending on
    the plasma model).

\item The diffuse X--ray emission at the center of NGC\,3077 is 
    hardened due to photoelectric absorption of cooler gas as
    traced by high--resolution \ion{H}{1} and CO observations. The
    diffuse emission is well described by collisional hot thermal
    plasma models.
    
\item A comparison of the diffuse X--ray emission with the
    H$\alpha$ morphology of NGC\,3077 reveals that regions
    containing hot gas are confined by expanding H$\alpha$ shells.
    The confinement is corroborated by the volume density profile
    which cannot be explained by a spherically--symmetric freely
    expanding wind.

\item For the individual superbubbles we derive the temperatures
    of the hot gas to $\sim 2\times 10^{6}$\,K, the masses to
    $\sim 10^{5}\,\xi^{-0.5}\,f_{v}^{0.5}$\,M$_{\sun}$, the
    particle volume densities to $\sim
    0.1\,\xi^{-0.5}\,f_{v}^{-0.5}$\,cm$^{-3}$, and the
    luminosities to $\sim 10^{38}$\,erg~s$^{-1}$, in agreement
    with theoretical models. The mass of the stellar ejecta and of
    material evaporated from the rim of the shells balances the
    mass of the hot gas within four shells. For two superbubbles,
    including that with the highest X--ray brightness, other
    mechanisms for mass--loading must be invoked. The high
    pressures ($P/k \sim
    10^{5-6}\,\xi^{-0.5}\,f_{v}^{-0.5}$\,K\,cm$^{-3}$) of the hot
    gas are believed to drive the expansion of the H$\alpha$
    shells. The entire diffuse X--ray luminosity is $\approx
    3.0\times 10^{39}$\,erg~s$^{-1}$; $\sim 1/3$ of the flux
    cannot be attributed to individual shells.\\

   \item The starburst properties of M\,82 and NGC\,3077 are quite
    different. In contrast to the superwind observed in M\,82, the
    hot gas in NGC\,3077 is still confined by the supershells.
    Also the physical parameters of the hot gas such thermal
    energy and mass of the hot gas as well as the current star
    formation rate are much higher in M\,82 in absolute terms
    (about two orders of magnitude). The temperature of the
    thermal plasma, however, is remarkably similar in M\,82,
    NGC\,3077, and in the non--starburst galaxy M\,81. The mass
    and thermal energy of the hot gas in M\,82 is still an order
    of magnitude higher compared to NGC\,3077 when normalizing
    these quantities to the galaxies' \ion{H}{1} masses or blue
    luminosities. Relative to their current star formation rates,
    however, both starbursts carry about the same amount of hot
    gas mass and thermal energy despite of the very different
    morphological and physical states of their starbursts.

   \item Spectral analysis of the six point sources reveal that
    NGC\,3077 hosts three young supernova remnants, two X--ray
    binaries, and one supersoft source. The first five sources are
    located close to the center of NGC\,3077, the latter one in
    the halo.

\item The X--ray source population in the tidal arm close to NGC\,3077
 is consistent with the extragalactic background population of the
 Chandra Deep Field South. An additional population of X--ray
 binaries or SNRs with fluxes $\gtrsim 6\times
 10^{-16}$\,erg~s$^{-1}$~cm$^{-2}$ (0.5--2\,keV) can be excluded.

\item Hot gas will likely escape from the gravitational potential of
 NGC\,3077 towards the north once the superbubbles reach sizes larger
 than the scalelength of the neutral gas ($\sim 0.5$\,kpc). To the
 south, however, the hot gas may not be able to overcome the pressure
 of the environmental, extended ISM and will return to its host. The
 gas driven out to the north will be stored in the intragroup medium of
 the M\,81 group.

\end{enumerate}

\acknowledgements JO acknowledges the Graduiertenkolleg on ``The
Magellanic System, Galaxy Interaction, and the Evolution of Dwarf
Galaxies" (GRK~118) of the ``Deutsche Forschungsgemeinschaft'' (DFG).
This work would not have been possible without the financial support
of SAO grant G01--2097X and the David and Lucile Packard Foundation
(CLM). We would like to thank Elias Brinks and the anonymous referee
for their valuable comments on the manuscript. This research has made
use of the NASA/IPAC Extragalactic Database (NED), which is maintained
by the Jet Propulsion Laboratory, Caltech, under contract with the
National Aeronautics and Space Administration (NASA), NASA's
Astrophysical Data System Abstract Service (ADS), NASA's SkyView, and
the astronomical database SIMBAD, provided by the ``Centre de
Donn\'ees astronomiques de Strasbourg'' (CDS).

\clearpage
\plotone{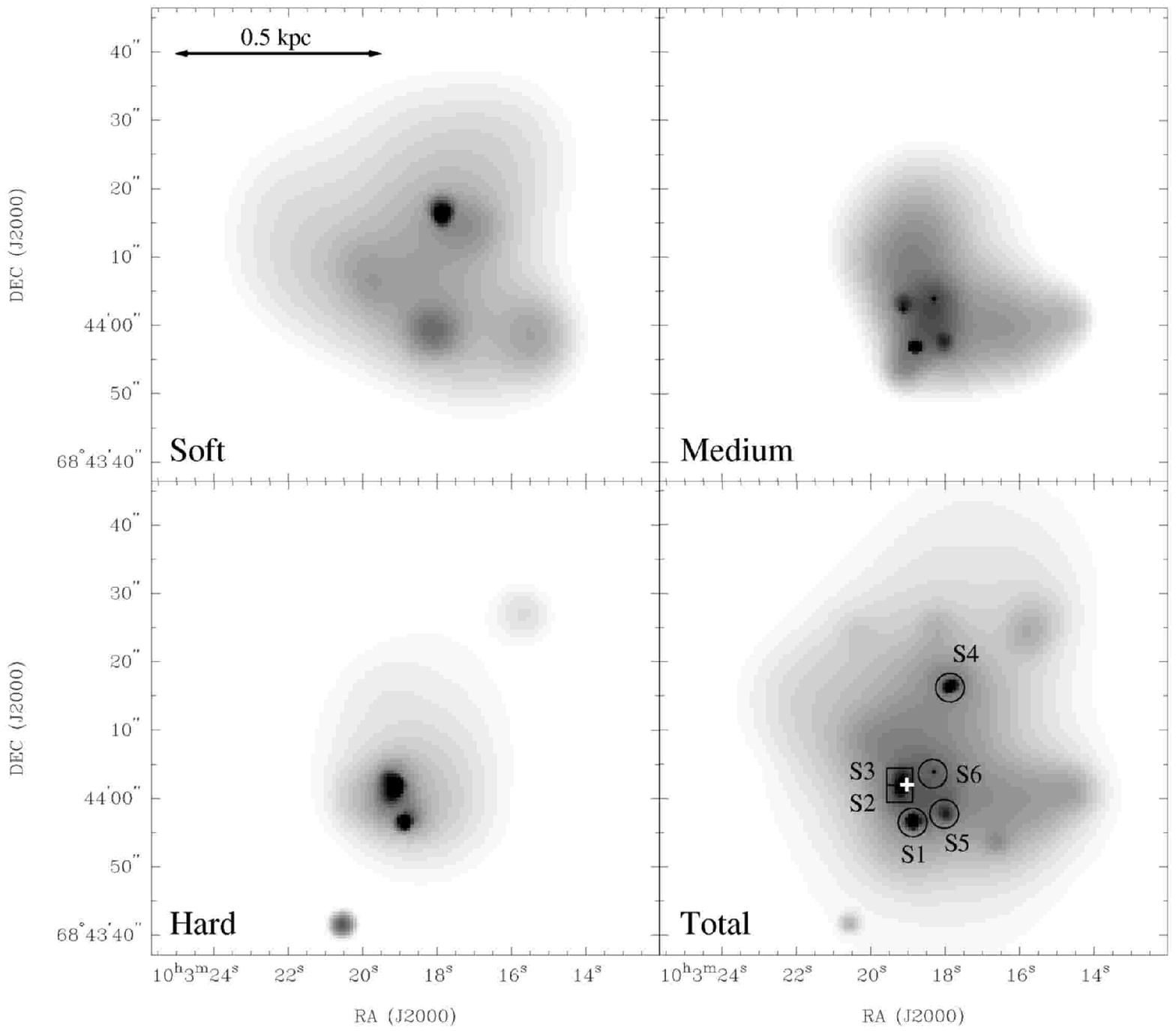}
\figcaption[f1.eps]
{X--ray emission of NGC\,3077 in different energy bands (logarithmic
scale). The data shown here are adaptively smoothed. {\bf Soft:}
0.3\,keV$\leqslant E \leqslant$0.7\,keV, {\bf Medium:}
0.7\,keV$\leqslant E \leqslant$1.1\,keV, {\bf Hard:}
1.1\,keV$\leqslant E \leqslant$6.0\,keV, and {\bf Total:}
0.3\,keV$\leqslant E \leqslant$6.0\,keV. The locations and definitions
of the point sources are labeled in the total band image and the white
{\bf plus} refers to the NICMOS H--band peak (see also
Fig.\,\ref{fig:color}).\label{fig:bands}}

\clearpage
    
\plotone{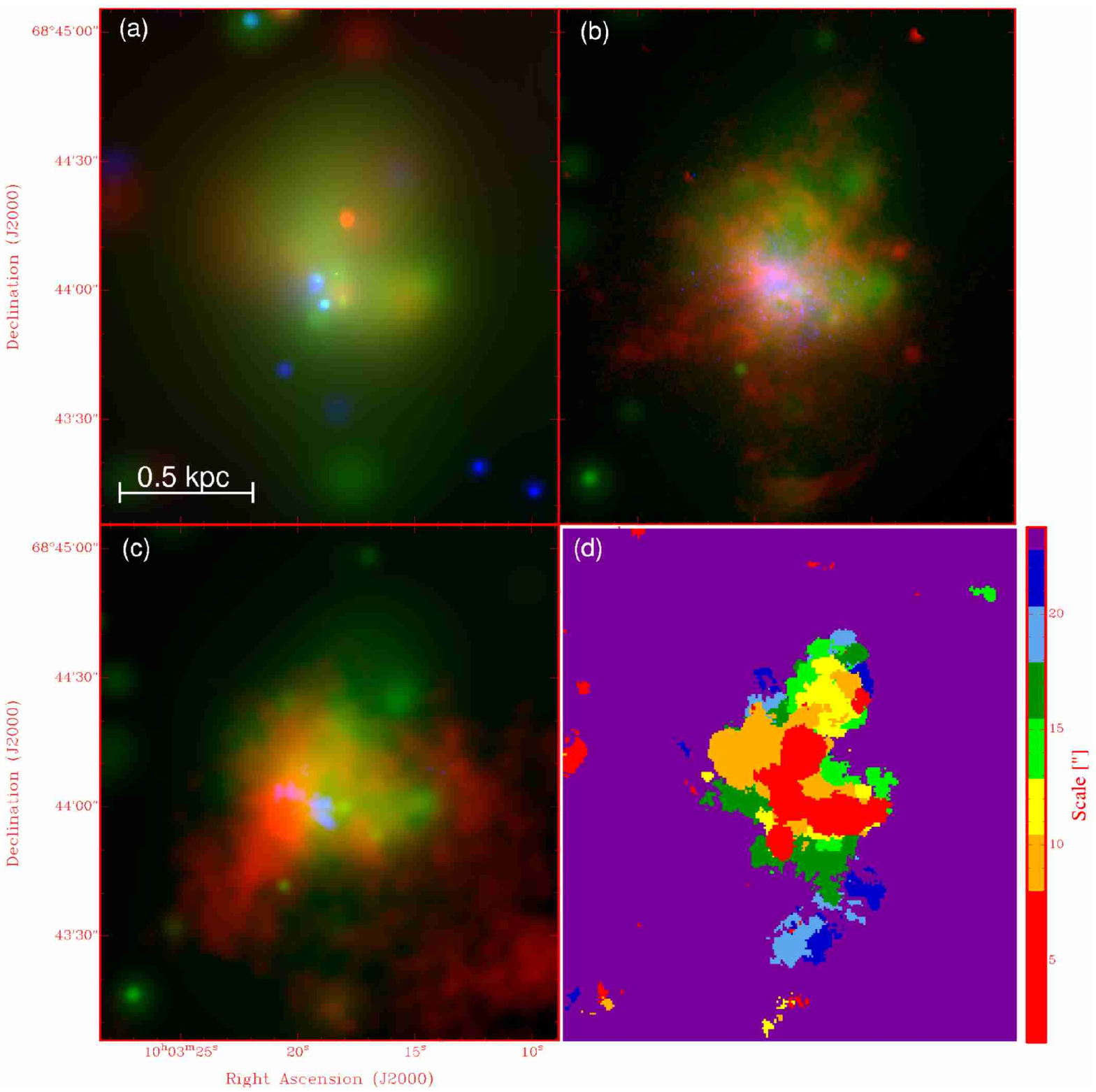} 
\figcaption[f2.eps] {{\bf a)}
 Three--color composite of the X--ray bands soft (red), medium
 (green), and hard (blue). {\bf b)} Three--color composite of
 H$\alpha$ (red), the total point source subtracted X--ray emission
 (green), and the H--band image (blue) of NGC\,3077. All images have
 been logarithmically scaled. Note that H$\alpha$ shells are filled
 with hot gas producing X--ray emission. This image forms the basis
 of the definition of the different regions in
 Fig.\,\ref{fig:regions}. {\bf c)} Three--color composite of the
 atomic hydrogen (\ion{H}{1}, red), the diffuse X--ray emission
 (green), and the CO content (blue). The X--ray emitting region is
 wedged between the cooler phases of the interstellar medium.
 1\arcmin\ corresponds to 1\,kpc. {\bf d)} The smoothing scales
 computed by the adaptive smoothing algorithm which were used to
 convolve the diffuse X--ray emission. All panels are on the same
 spatial scale.
 \label{fig:color}} 
\epsscale{1}

\clearpage
\epsscale{0.8}
\plotone{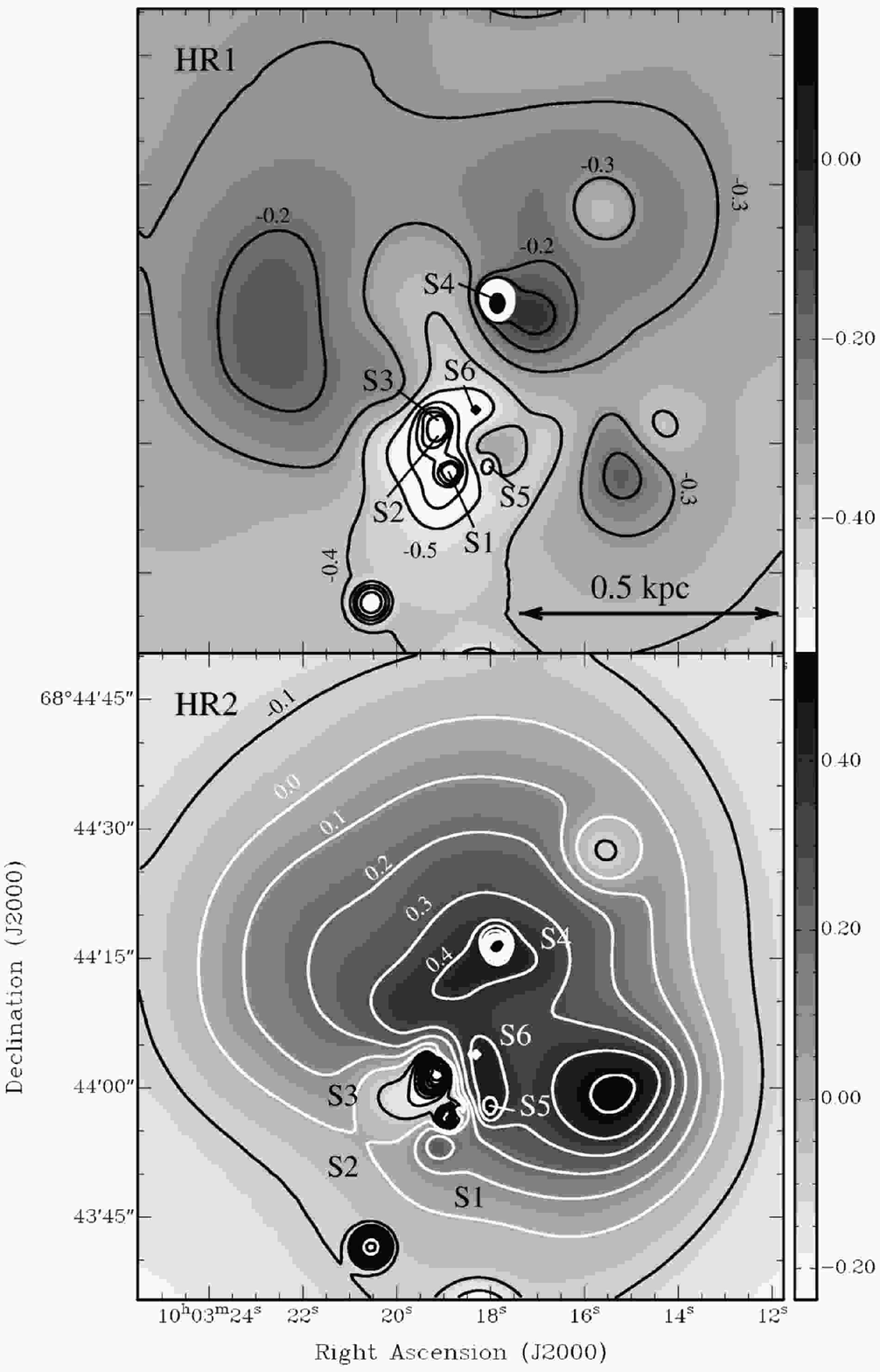}
\figcaption[f3.eps]
{Hardness ratio images of NGC\,3077 using the normalized adaptively
smoothed X--ray images. The ratios are defined as
HR1=(Soft$-$Medium$-$Hard)/(Soft+Medium+Hard) ({\bf top}) and
HR2=(Soft+Medium$-$Hard)/(Soft+Medium+Hard) ({\bf bottom}). Both
images show the same region. The contours change their color at HR=0
and they are separated by 0.1. Point sources are labeled according to
Fig.\,\ref{fig:bands}.\label{fig:hr_images}}
\epsscale{1}

\clearpage 
\plotone{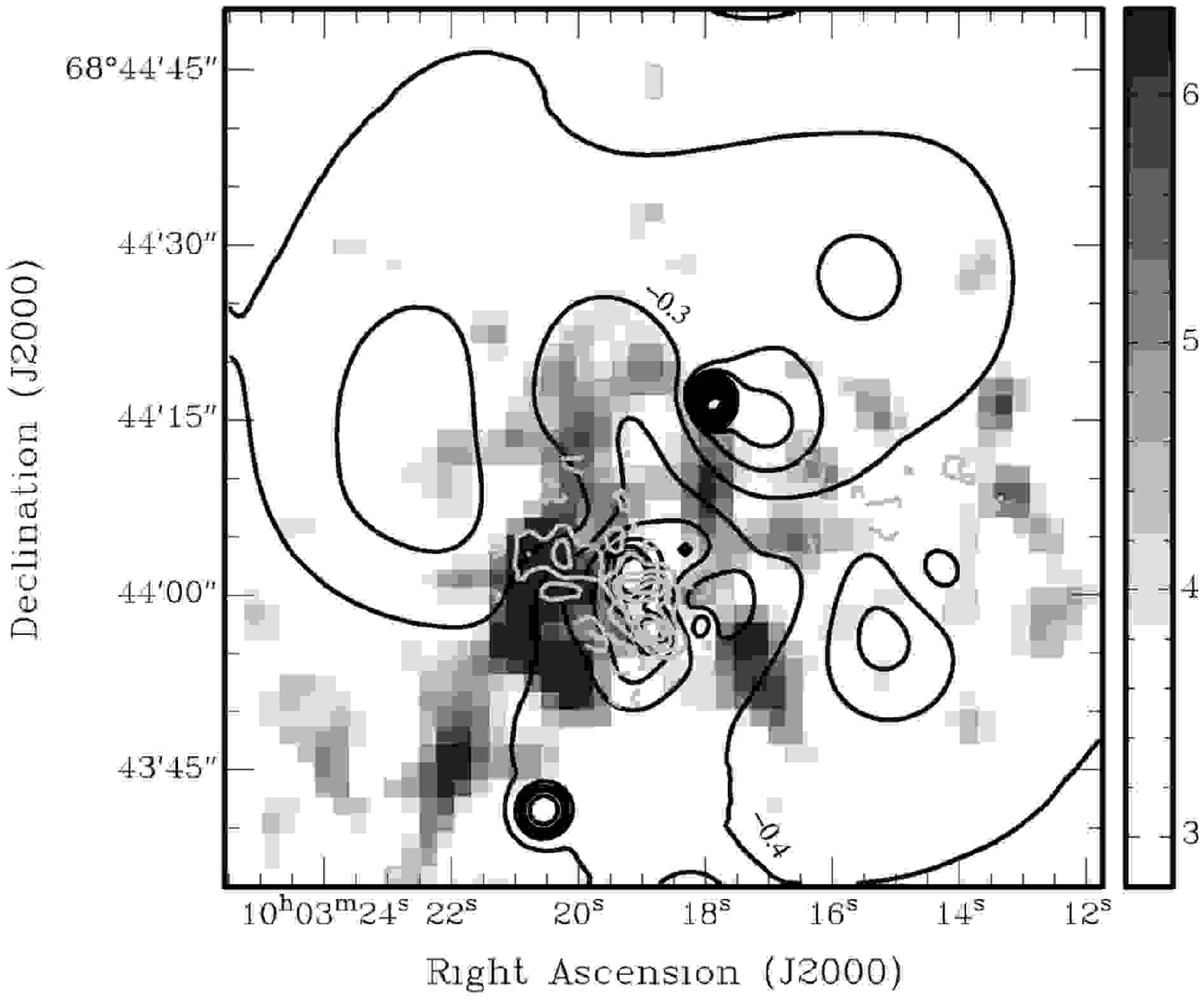} 
\figcaption[f4.eps] {{\bf Grey scale}:
 High--resolution (6\farcs5) \ion{H}{1} column density map of
 NGC\,3077 in units of 10$^{21}$\,cm$^{-2}$. {\bf Black contours}:
 The X--ray hardness ratio image HR1 (same contour levels as in
 Fig.\,\ref{fig:hr_images}). {\bf Grey contours}: CO (1--0) line
 emission (CO data taken from W02). 
\label{fig:hard_hi}}

\clearpage
\plotone{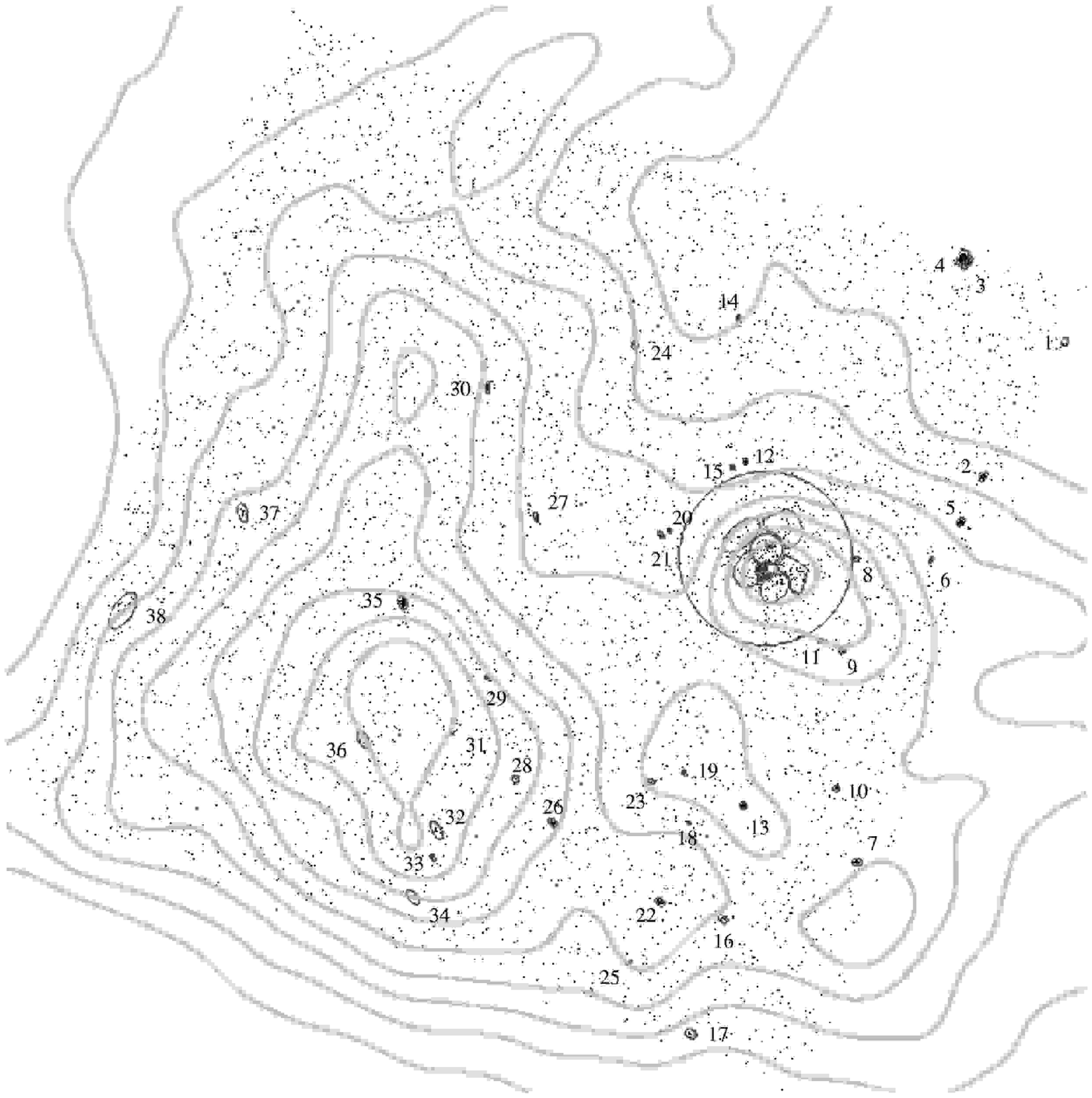}
\figcaption[f5.eps]
{The raw ACIS--S3 data superposed with \ion{H}{1} contours of the
tidal tail. The contours start at a column density of $2.76\times
10^{20}$\,cm$^{-2}$ and are equally spaced by the same amount
(resolution: 40\arcsec). Point sources detected by the wavelet
algorithm ({\it wavdetect}) are marked with small ellipses. The numbers
correspond to those in Table\,\ref{tab:field}. Additionally, we
show the definition of the diffuse emitting regions evaluated in
Table\,\ref{tab:regions}.\label{fig:field}}

\clearpage
\plotone{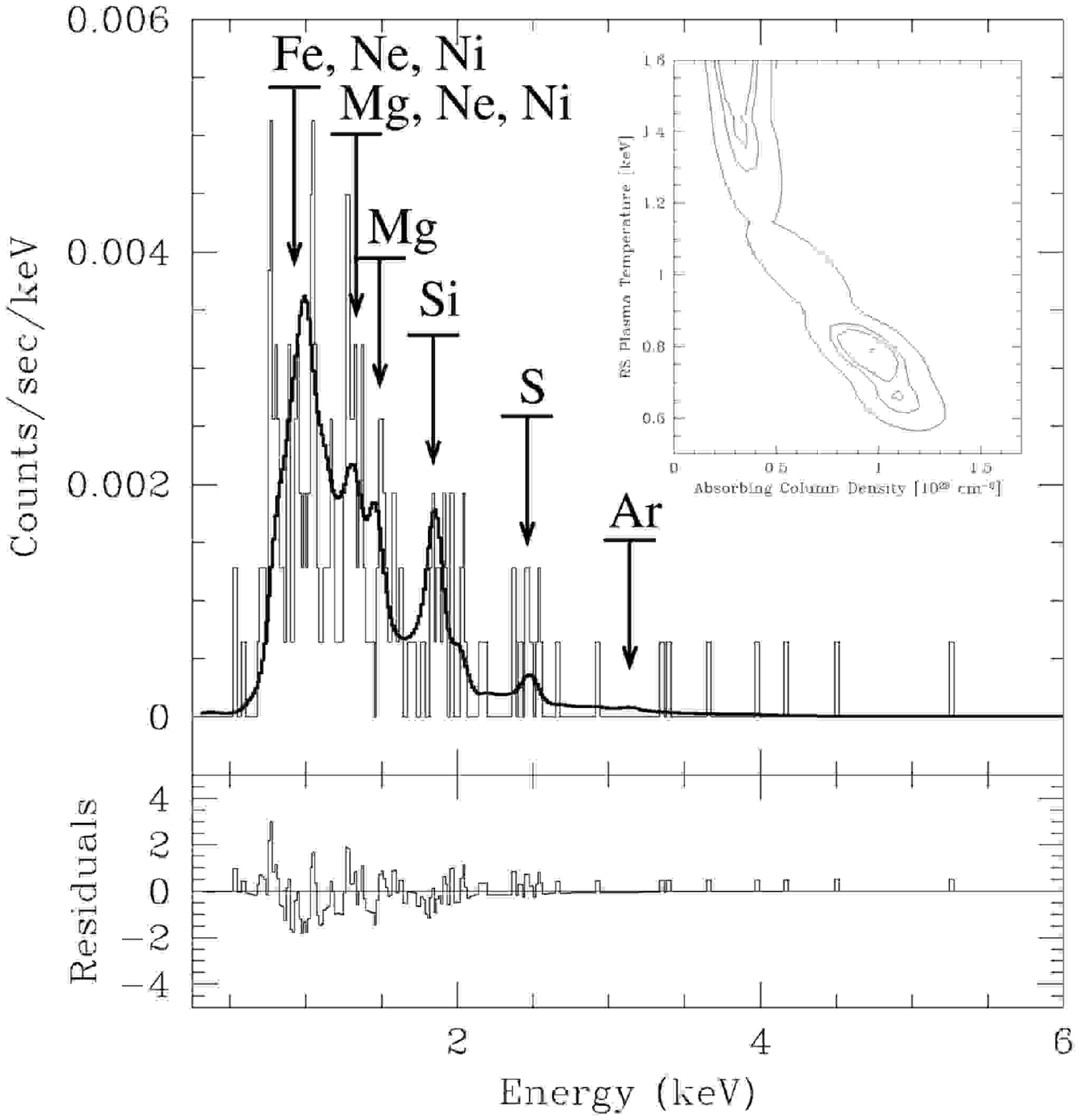}
\figcaption[f6.eps]
{The ACIS--S3 spectrum of the point source S1. The best fitting
RS spectrum is overlaid and the residuals ({\it
Data$-$Model} in counts) of the fit are shown in the lower panel. We
marked the elements responsible for major line complexes. The {\bf inset}
displays the confidence regions in the $N_{H}-T$ plane (68.3\%,
90.0\%, and 99.0\% confidence levels; the best fit is marked by the
{\bf plus}). \label{fig:p1}}

\clearpage 
\plotone{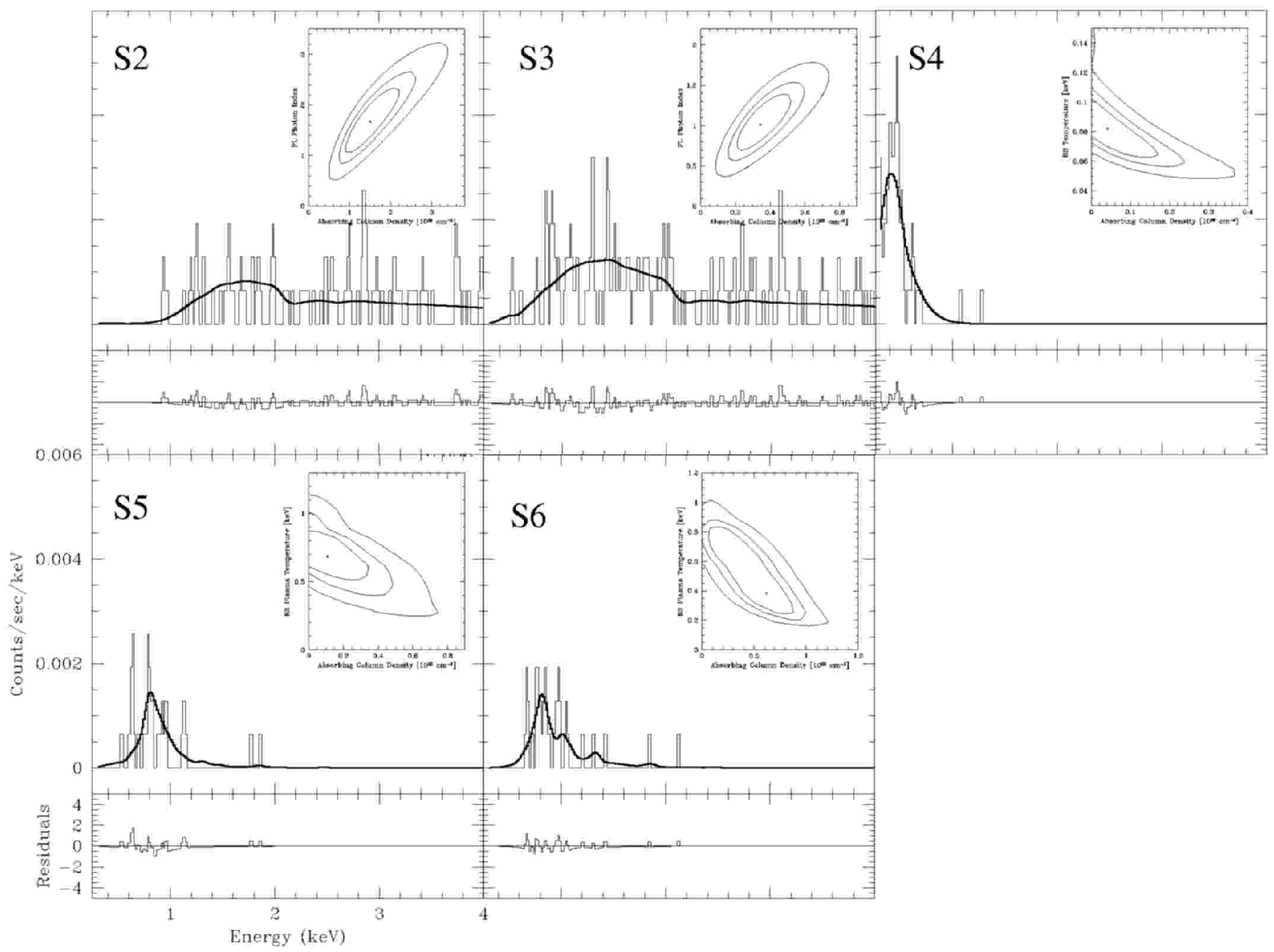} 
\figcaption[f7.eps] {ACIS--S3 spectra of
 the point sources S2--S6. Overlaid are the best fits for the
 following models. {\bf S2} and {\bf S3}: power law, {\bf S5} and
 {\bf S6}: RS, {\bf S4}: black body (see
 Table\,\ref{tab:pointsources}). The residuals of these fits ({\it
  Data$-$Model} in counts) are shown in the lower panels. The insets
 display $N_{H}-T$ confidence regions at 68.3\%, 90.0\%, and 99.0\%
 levels; the {\bf plus} indicates the best fit. Panels showing
 spectra and residuals are all on the same scale. \label{fig:p23456} }

\clearpage
\plotone{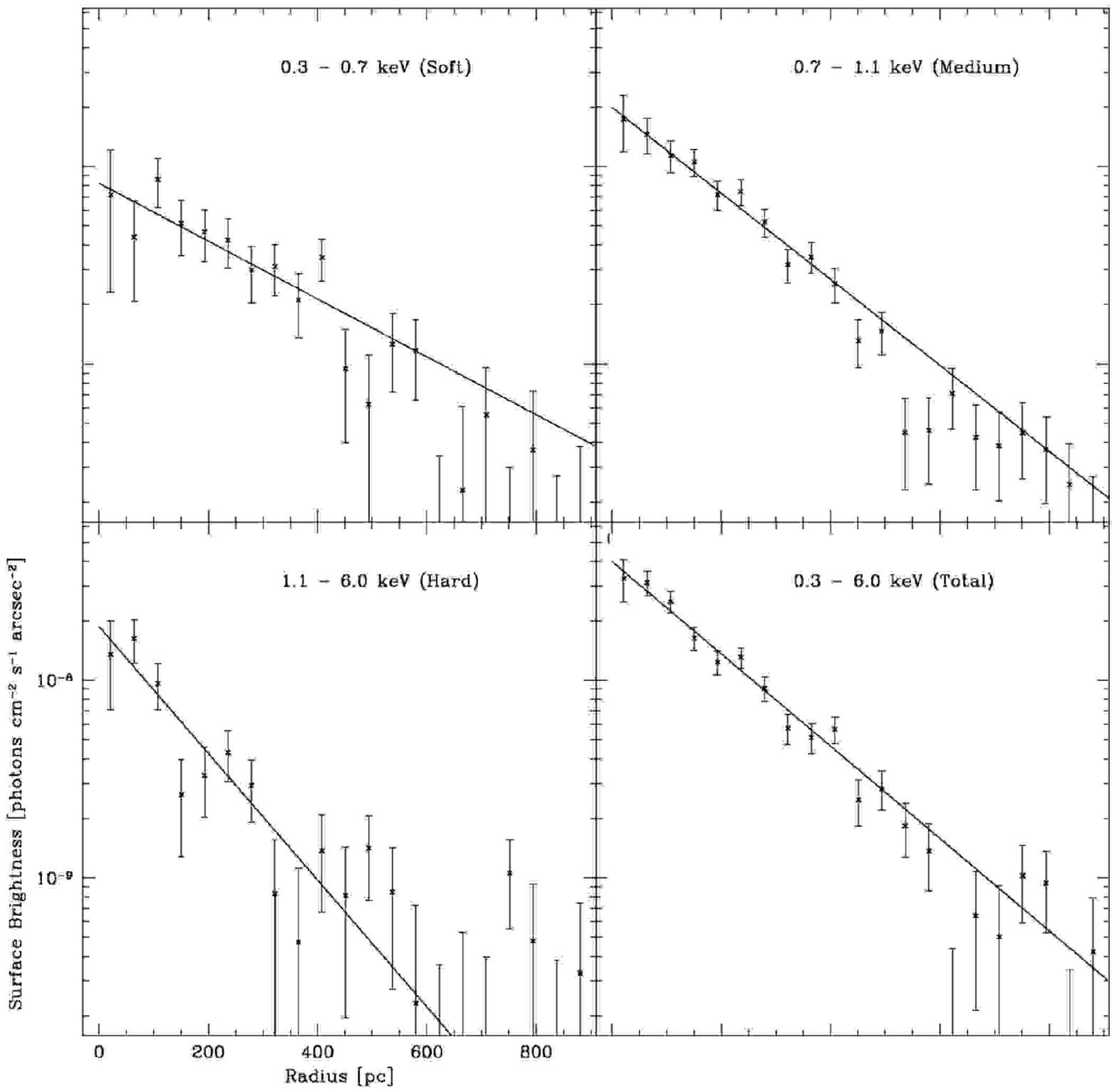}
\figcaption[f8.eps]
{Azimuthally--averaged X--ray surface brightness profiles for different energy
ranges, all on the same logarithmic scale. The curves are best exponential
fits for radii $< 700$\,pc. \label{fig:surflog}}

\clearpage
\plotone{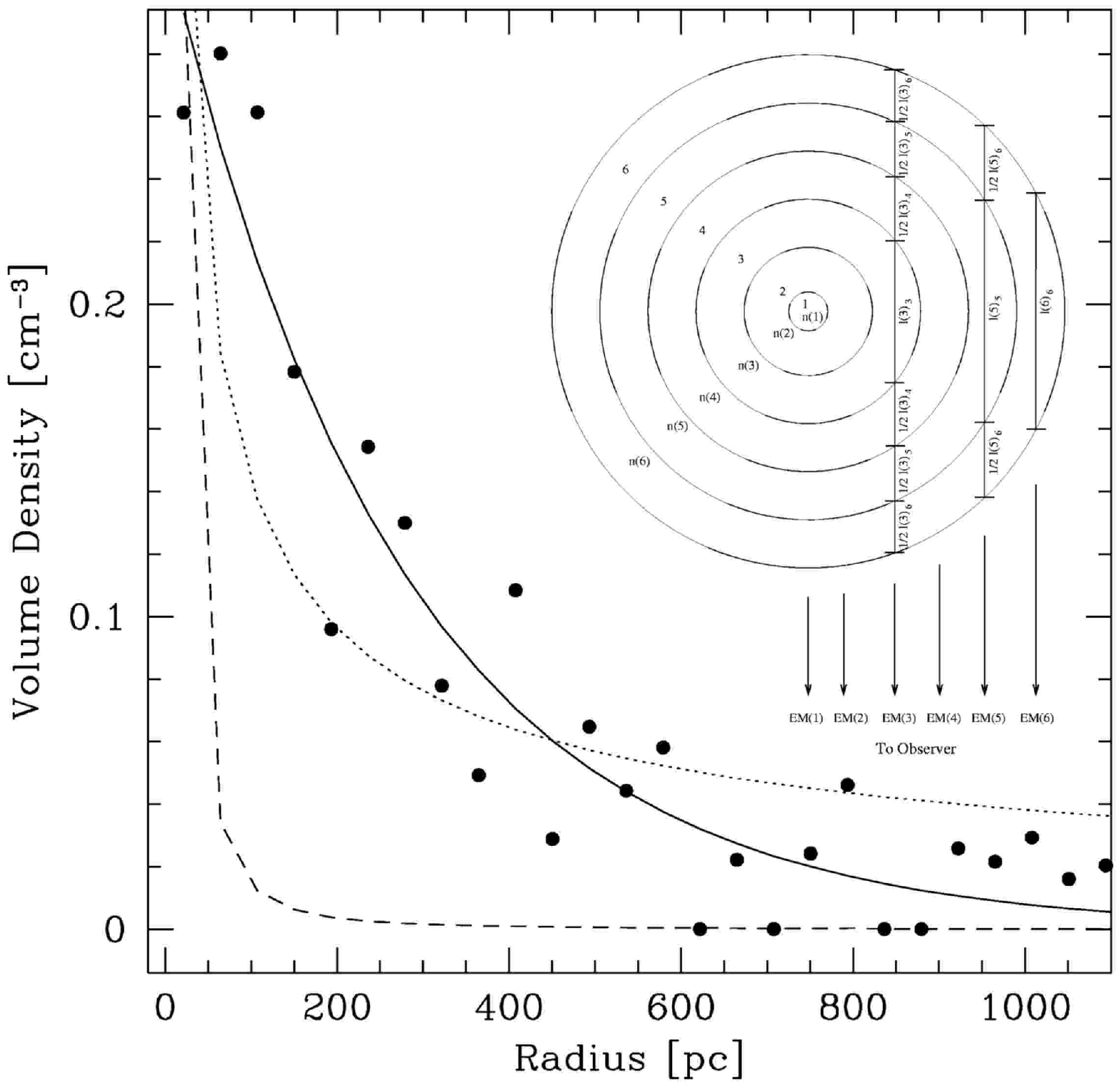}
\figcaption[f9.eps]
{Radial profile of the three dimensional volume density distribution
of the hot gas. Best fits of the following functions are
overlaid. {\bf Solid line}: exponential, {\bf dotted line}: power law,
{\bf dashed line}: power law with an index of $\beta=2$. The inset of
this figure is an example of how a profile using six shells is
calculated (see Sect.\,\ref{sec:diffuse_reduc} for
details).\label{fig:density}}

\clearpage 
\epsscale{0.5} 
\plotone{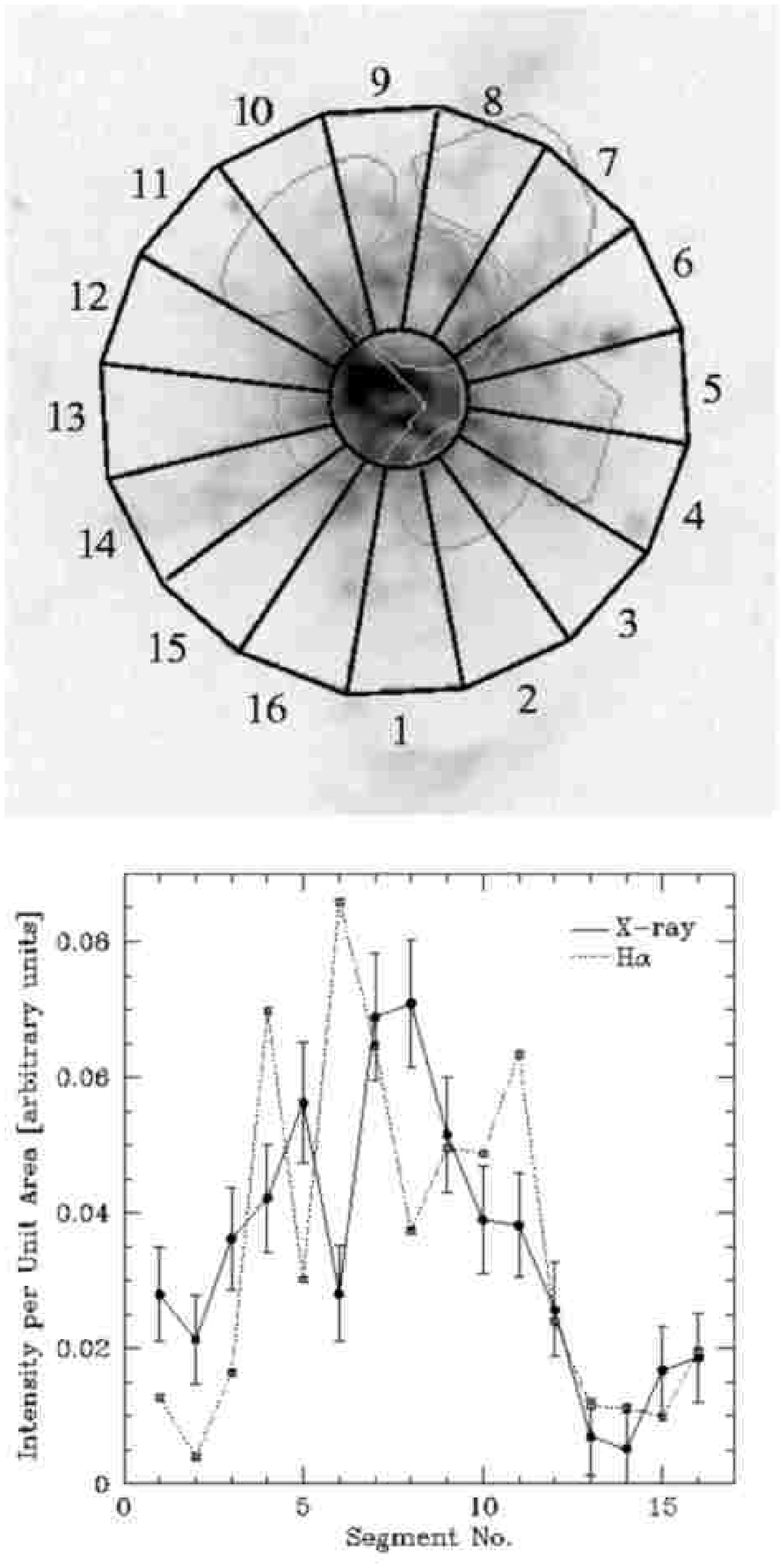} 
\figcaption[f10.eps]
{Azimuthal dependence of the X--ray and H$\alpha$ features in
 NGC\,3077. The {\bf upper panel} shows the numbering of the
 individual segments displayed as {\bf black polygons}. For
 illustration, the regions defined in Fig.\,\ref{fig:regions} are
 overlaid in {\bf grey}. {\bf Lower panel}: The intensity per unit
 area of the total X--ray band and the H$\alpha$ image within the
 corresponding segments in arbitrary units.
 \label{fig:cake}}

\epsscale{1}

\clearpage
\plotone{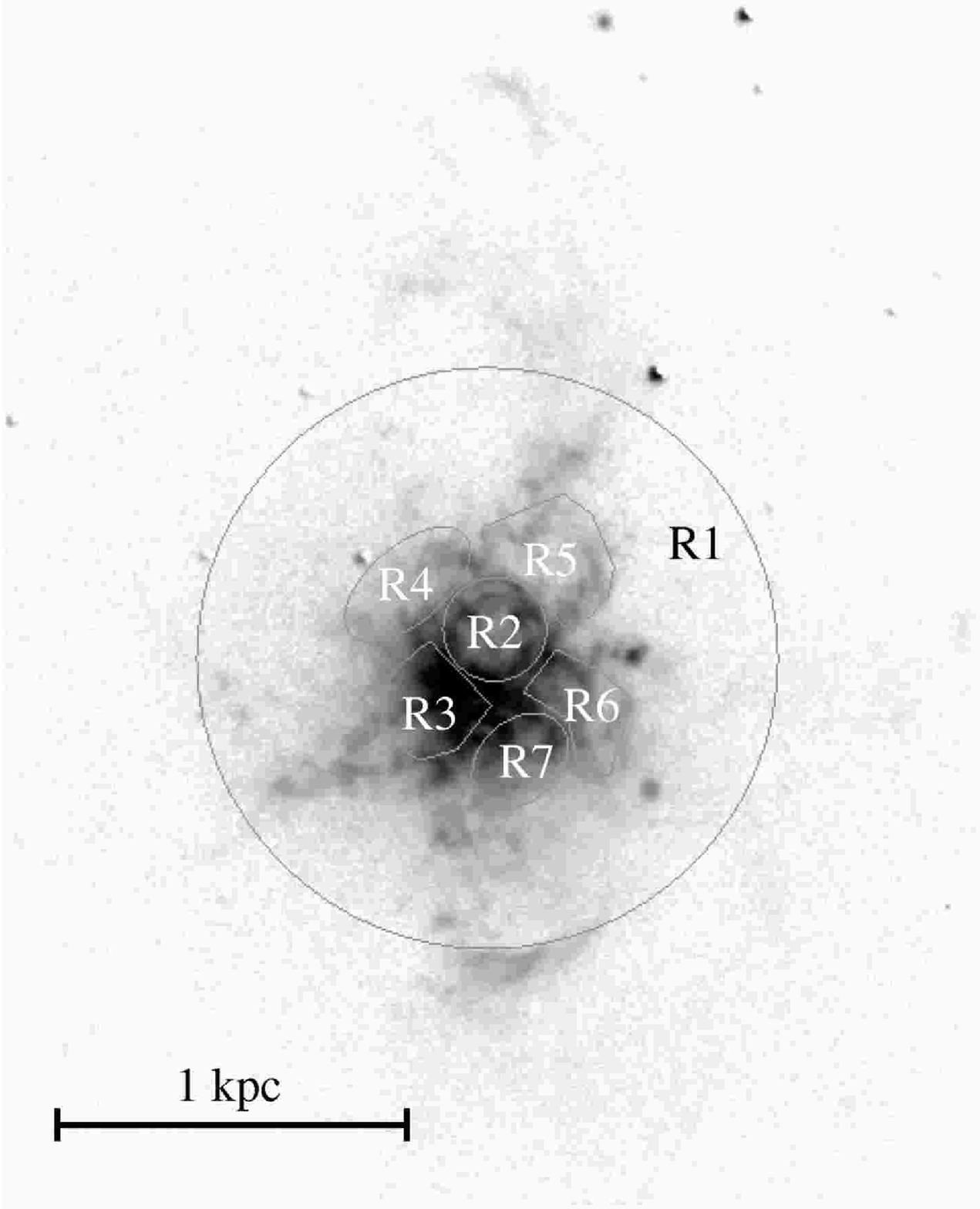}
\figcaption[f11.eps]
{Definition of the regions R1 to R7 overlaid on an H$\alpha$
image.\label{fig:regions}}

\clearpage
\plotone{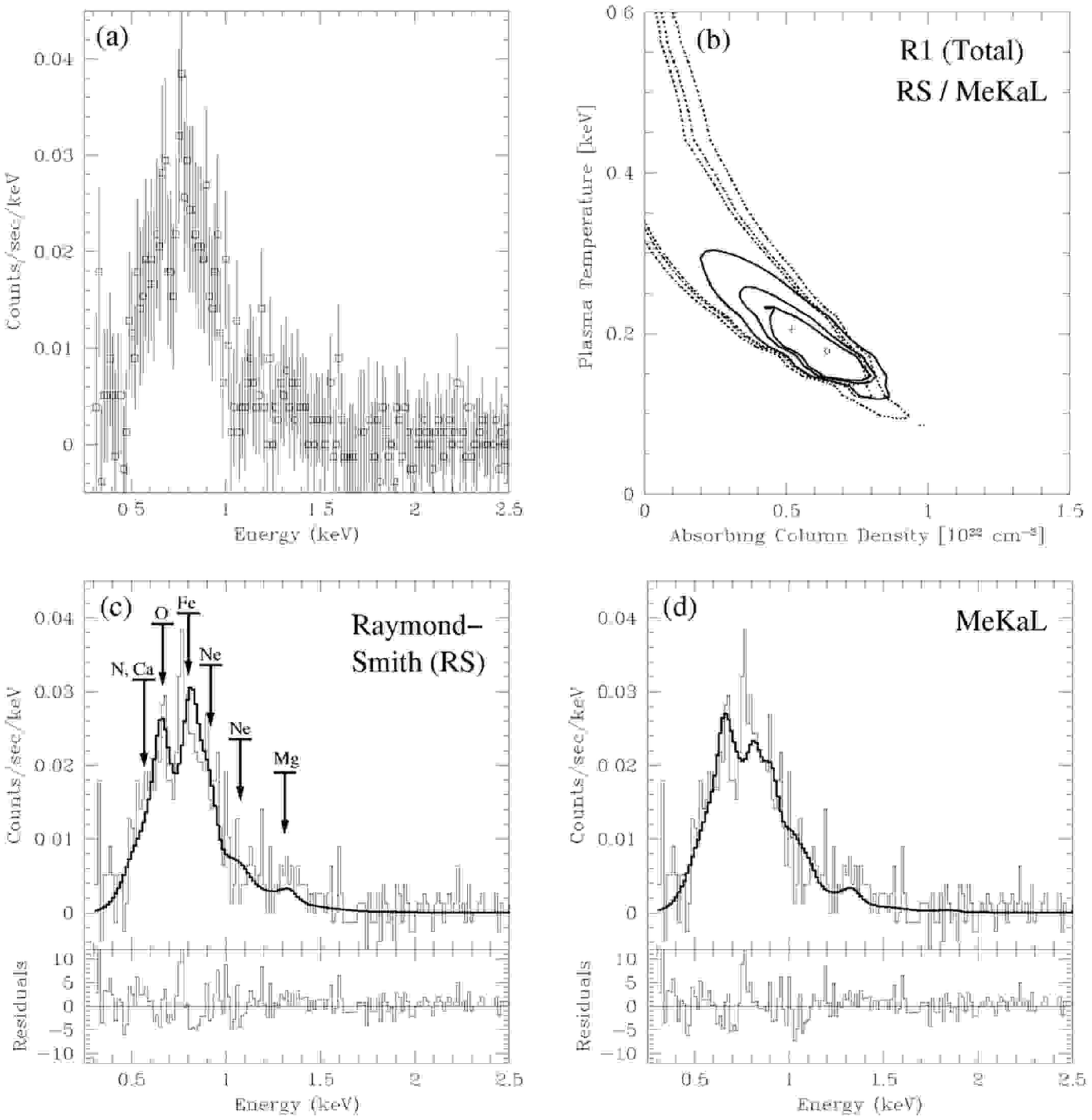}
\figcaption[f12.eps]
{{\bf a)} The background--subtracted spectrum of the total diffuse
emission of NGC\,3077 (R1). Identified point sources were previously
removed. {\bf b)} Confidence regions for the two different plasma
models (RS and MeKaL) in 68.3\%, 90.0\%, and 99.0\%
intervals. {\bf Solid}: RS model (best fit visualized by a
circle), {\bf Dotted}: MeKaL plasma (plus). The corresponding best
fitting spectra of the models are overlaid to the measured spectrum in
panels {\bf c)} and {\bf d)}, as well as the elements responsible for
prominent line complexes.
\label{fig:spectra}}

\clearpage 
\epsscale{0.8} 
\plotone{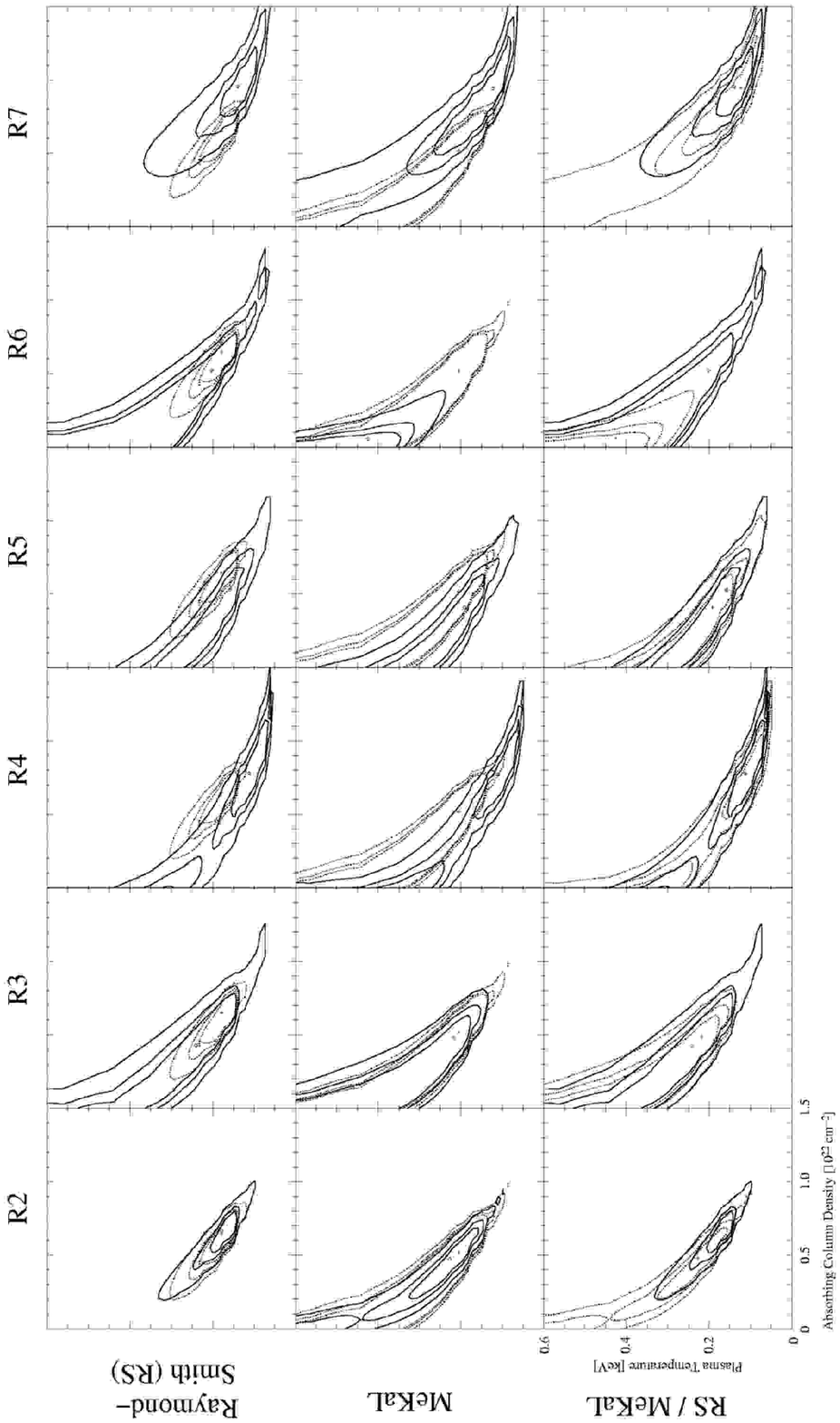} 

\figcaption[f13.eps]
{$N_{H}-T$ Confidence regions (68.3\%, 90.0\%, and 99.0\%) of RS and
 MeKaL plasma models for the individual regions R2 to R7. The {\bf
  plus} ({\bf circle}) refers to the best fit of the model shown in
 {\bf solid} ({\bf dotted}) contours. All plots are on the same
 scale. {\bf First row}: Both sets of contours are from RS plasma
 models. The {\bf dotted} ones are for the overall diffuse emission
 of R1, the {\bf solid} ones for the individual bubbles. {\bf Second
  row}: Like the first row, but for the MeKaL models. {\bf Third
  row}: Overlay of the MeKaL models ({\bf dotted}) on the RS models
 ({\bf solid}) for each region (for the R1, the diffuse emission of
 the entire galaxy, see Fig.\,\ref{fig:spectra}).
\label{fig:contourplots}}
\epsscale{1}

\clearpage 
\plotone{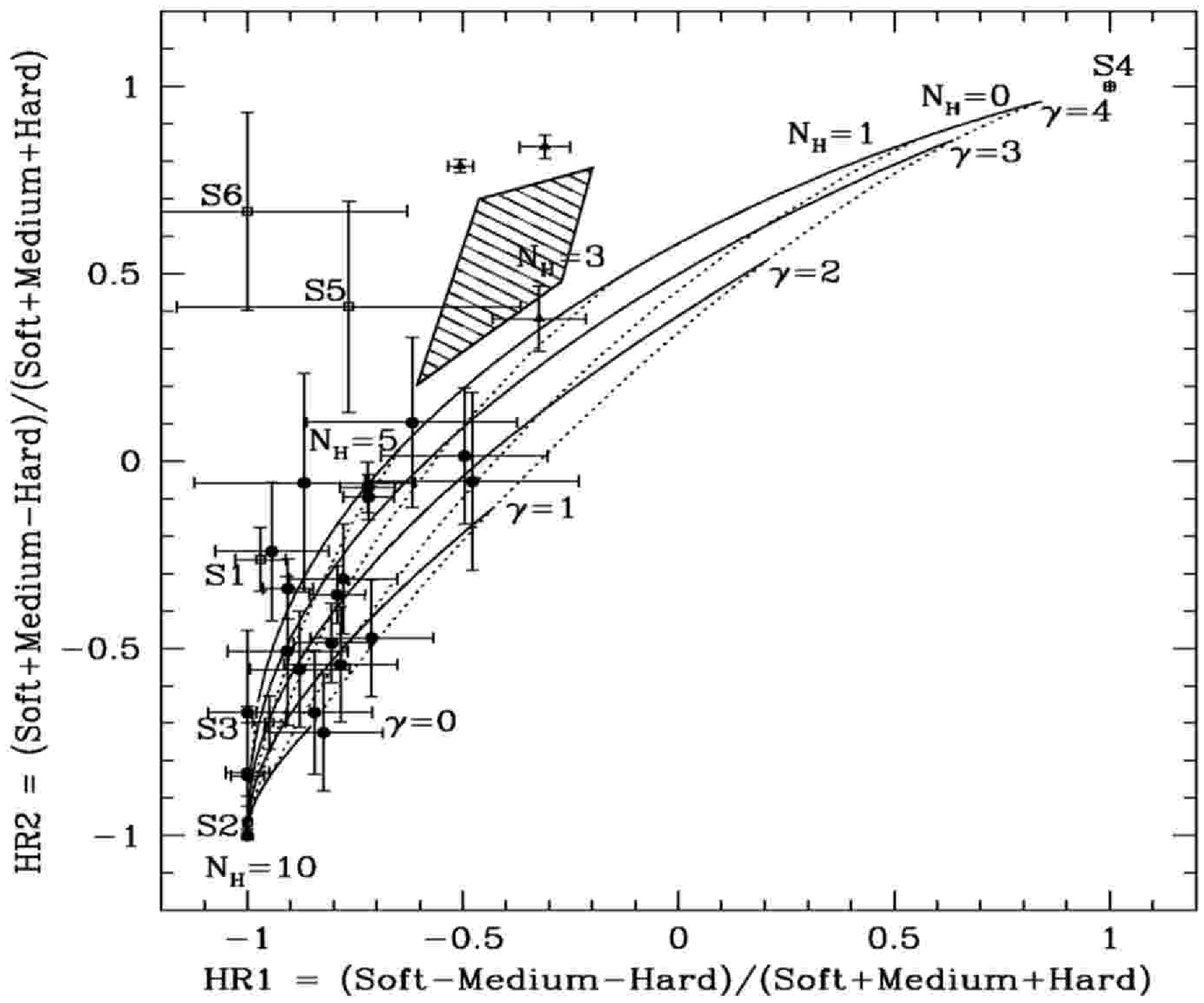} 

\figcaption[f14.eps] {Hardness ratio plot of the point sources in the
 Chandra ACIS--S3 field of view which were detected by both {\it
  wavdetect} and {\it celldetect} source detection algorithms. {\bf
  Open Squares:} The point sources within the diffuse X--ray
 emission of NGC\,3077. The sources are marked according to
 Fig.\,\ref{fig:bands} and Table\,\ref{tab:pointsources}. {\bf Filled
  Circles:} Background Sources. Objects which are identified as
 stars are plotted as {\bf filled triangles}. Power law models are
 overlaid on the data. {\bf Solid lines:} The photon indices $\gamma$
 are kept fixed, while the absorbing column densities vary from 0 to
 10$^{22}$\,cm$^{-2}$. {\bf Dotted lines:} For a fixed absorbing
 column density $N_{H}$ the photon indices change from 0 to 4. Column
 densities are given in units of 10$^{21}$\,cm$^{-2}$. The {\bf
  shaded region} marks the positions of the regions R1 to R7.
 Thermal plasma models with temperatures below $\sim 15\times
 10^{6}$\,K occupy the region in the upper left corner -- well above
 the power law models.
\label{fig:hardness}}

\clearpage
\epsscale{0.6}
\plotone{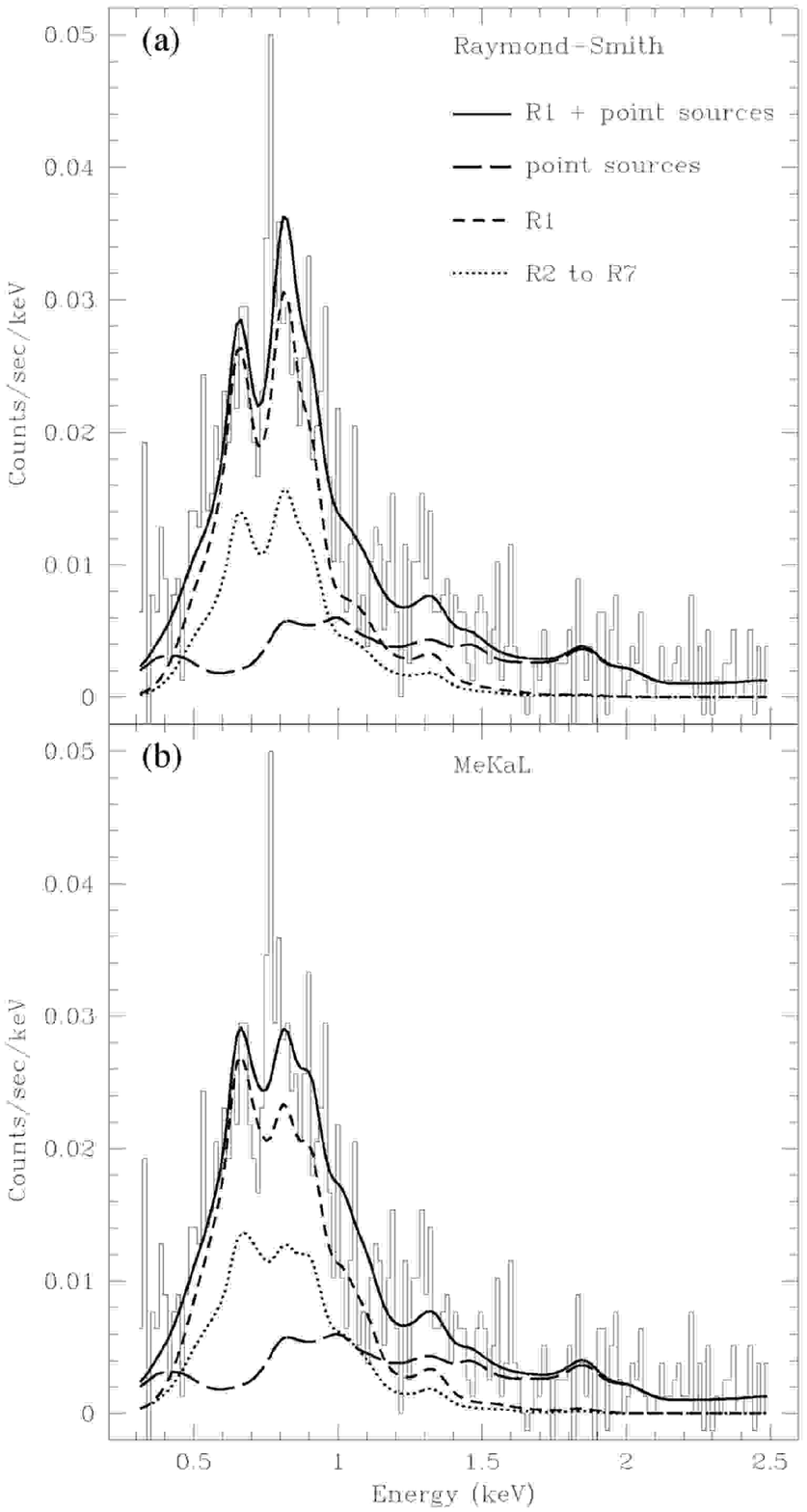}
\figcaption[f15.eps]
{Total X--ray spectrum of NGC\,3077. Overlaid are the best fits to the point
sources and to the diffuse emission regions, the latter modeled by
{\bf a}): RS models and {\bf b}): MeKaL models. R1
corresponds to the total diffuse X--ray emission of NGC\,3077.
\label{fig:tot_spectrum}}
\epsscale{1}

\clearpage
\epsscale{0.6}
\plotone{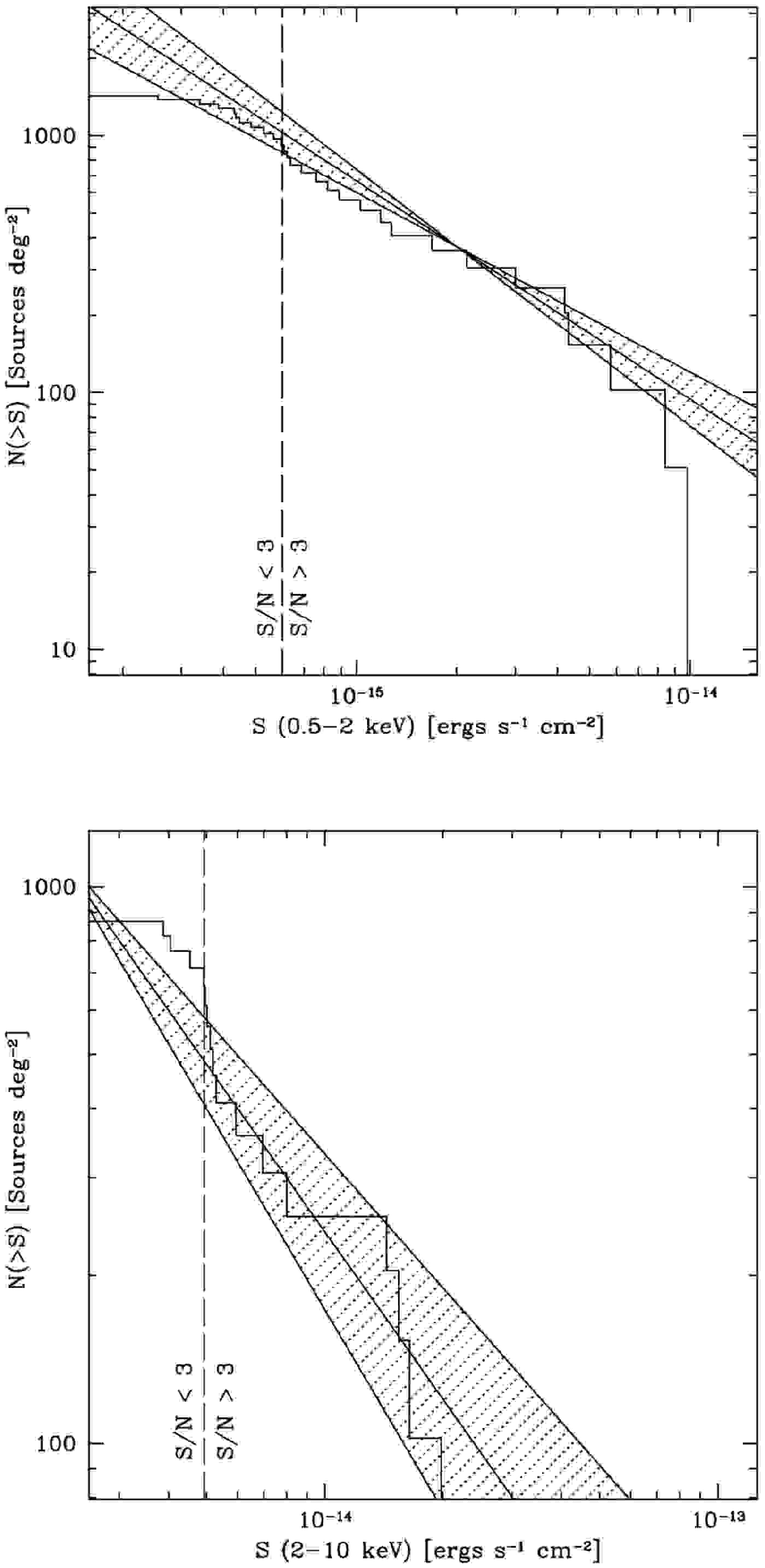}

\figcaption[f16.eps] {$Log N(>S)$--$Log S$ plots of point sources in
 the G01 soft and hard bands within the ACIS--S3 field of view
 (binned to 1\arcsec\ pixel size). We excluded sources found within
 the hot outflow of NGC\,3077 as well as optically identified stars
 (corresponding to objects 3, 4, and 7 in Fig.\,\ref{tab:field} and
 Table\,\ref{tab:field}). The data are compared to the results from
 G01 for the CDFS ({\bf solid lines}), where the {\bf shaded regions}
 mark the error of their fit. The limiting fluxes ($S/N<3$) are
 indicated by {\bf dashed lines}.\label{fig:logn}} \epsscale{1}

\clearpage
\plotone{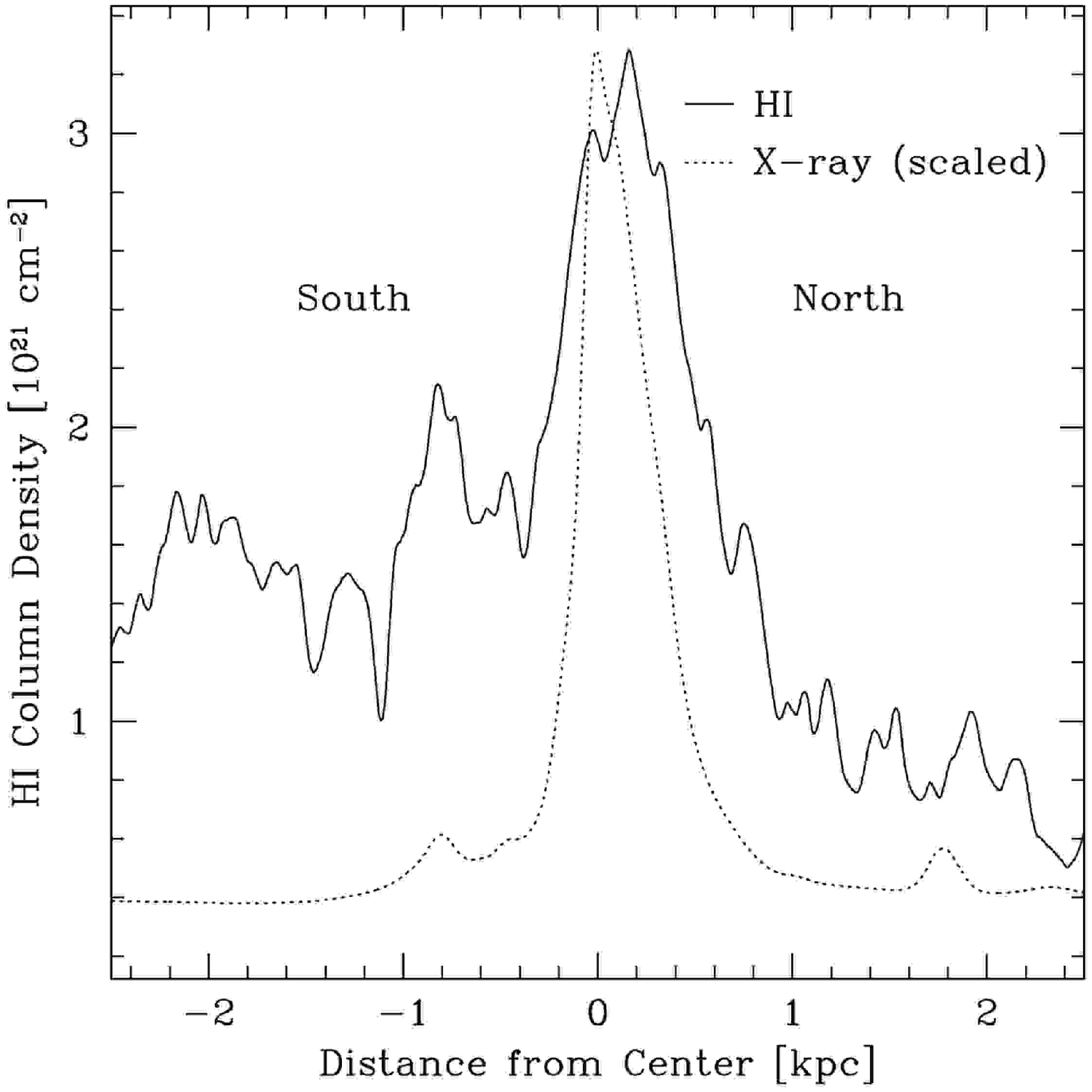}
\figcaption[f17.eps]
{\ion{H}{1} column density profile along the north--south direction of
NGC\,3077 through the H--band peak (the center of NGC\,3077). For a
comparison, we also show a profile of the diffuse X--ray emission
along the same direction. The latter has been scaled to match the
\ion{H}{1} peak intensity. \label{fig:hi_profile}}

\clearpage
\begin{deluxetable}{rcrrrrrrrr}

\tablecolumns{10}
\tabletypesize{\scriptsize}
\rotate
\tablewidth{0pt} 
\tablecaption{Source list of the ACIS--S3 point sources (exposure time: 53.4\,ks).}

\tablehead{
No.&\multicolumn{1}{c}{Algorithm}&\multicolumn{1}{c}{RA (J2000)\tablenotemark{a}~~~}&\multicolumn{1}{c}{~~~DEC (J2000)\tablenotemark{a}}&\multicolumn{1}{c}{Total Counts}&\multicolumn{1}{c}{Soft Counts}&\multicolumn{1}{c}{Medium Counts}&\multicolumn{1}{c}{Hard Counts}&\multicolumn{1}{c}{HR1}&\multicolumn{1}{c}{HR2}\\
\multicolumn{1}{c}{(1)}&\multicolumn{1}{c}{(2)}&\multicolumn{1}{c}{(3)}&\multicolumn{1}{c}{(4)}&\multicolumn{1}{c}{(5)}&\multicolumn{1}{c}{(6)}&\multicolumn{1}{c}{(7)}&\multicolumn{1}{c}{(8)}&\multicolumn{1}{c}{(9)}&\multicolumn{1}{c}{(10)}}

\startdata
 $^*1$ &w&  10 \, 02 \, 45.2 &  68 \, 46 \, 18.9 &$   11.0 \pm   3.5     $&$        0.0  \pm   0.0      $&$         0.0  \pm    0.0   $&$      11.4  \pm     3.5$&$-1.00\pm 0.00$&$  -1.00 \pm 0.00$\\                              
 2 &w/c&  10 \, 02 \, 54.2 &  68 \, 44 \, 57.5 &$   68.5 \pm   8.4     $&$        6.7  \pm   2.6      $&$        10.9  \pm    3.3   $&$      50.9  \pm     7.2$&$-0.80\pm 0.09$&$                             -0.49 \pm 0.11$\\  
 $^*3$\tablenotemark{b} &w/c&  10 \, 02 \, 56.0 &  68 \, 47 \, 06.9 &$  333.8 \pm  18.4     $&$      115.1  \pm  10.8      $&$       191.8  \pm   13.9   $&$      26.9  \pm     5.4$&$-0.31\pm 0.06$&$        +0.84 \pm 0.03$\\  
 $^*4$\tablenotemark{c} &w/c&  10 \, 02 \, 56.4 &  68 \, 47 \, 09.5 &$ 1273.6 \pm  35.7     $&$      314.3  \pm  17.7      $&$       823.7  \pm   28.7   $&$     135.5  \pm    11.8$&$-0.51\pm 0.03$&$        +0.79 \pm 0.02$\\     
 5 &w/c&  10 \, 02 \, 56.7 &  68 \, 44 \, 30.5 &$  140.5 \pm  11.9     $&$        6.6  \pm   2.7      $&$        39.7  \pm    6.3   $&$      94.2  \pm     9.7$&$-0.91\pm 0.06$&$                             -0.34 \pm 0.08$\\  
$^*6$ &w&  10 \, 03 \, 00.0 &  68 \, 44 \, 07.8 &$    5.6 \pm   2.5     $&$        0.0  \pm   0.0      $&$         0.0  \pm    0.0   $&$       5.7  \pm     2.5$&$-1.00\pm 0.00$&$                            -1.00 \pm 0.00$\\ 
$^*7$ &w/c&  10 \, 03 \, 08.2 &  68 \, 41 \, 05.8 &$  116.4 \pm  10.9     $&$       39.4  \pm   6.3      $&$        40.9  \pm    6.4   $&$      36.1  \pm     6.1$&$-0.32\pm 0.11$&$          +0.38 \pm 0.09$\\ 
 8 &w/c&  10 \, 03 \, 08.4 &  68 \, 44 \, 08.6 &$   28.2 \pm   5.4     $&$        0.8  \pm   1.0      $&$         9.9  \pm    3.2   $&$      17.5  \pm     4.2$&$-0.94\pm 0.13$&$                             -0.24 \pm 0.19$\\   
 9 &w/c&  10 \, 03 \, 09.9 &  68 \, 43 \, 13.3 &$   14.2 \pm   3.9     $&$        0.0  \pm   0.0      $&$         0.0  \pm    0.0   $&$      14.5  \pm     3.9$&$-1.00\pm 0.00$&$                             -1.00 \pm  0.00$\\                            
10\tablenotemark{d} &w/c&  10 \, 03 \, 10.5 &  68 \, 41 \, 50.2 &$   43.2 \pm   6.6     $&$        4.8  \pm   2.2      $&$        10.0  \pm    3.2   $&$      28.4  \pm     5.4$&$-0.78\pm 0.13$&$            -0.31 \pm 0.15$\\  
11 &w/c&  10 \, 03 \, 12.3 &  68 \, 43 \, 18.9 &$   10.5 \pm   3.3     $&$        0.0  \pm   0.0      $&$         0.9  \pm    1.0   $&$       9.8  \pm     3.2$&$-1.00\pm 0.05$&$                             -0.83 \pm 0.18$\\    
12 &w/c&  10 \, 03 \, 20.6 &  68 \, 45 \, 07.1 &$   19.3 \pm   4.5     $&$        3.7  \pm   2.0      $&$         7.0  \pm    2.6   $&$       8.7  \pm     3.0$&$-0.62\pm 0.24$&$                             +0.10 \pm 0.23$\\  
$^*13$ &w/c&  10 \, 03 \, 20.8 &  68 \, 41 \, 40.0 &$  221.0 \pm  14.9     $&$       30.8  \pm   5.6      $&$        71.9  \pm    8.5   $&$     118.3  \pm    10.9$&$-0.72\pm 0.06$&$                         -0.07 \pm  0.07$\\
$^*14$ &w&  10 \, 03 \, 21.4 &  68 \, 46 \, 33.4 &$    6.5 \pm   2.6     $&$        1.8  \pm   1.4      $&$         2.0  \pm    1.4   $&$       2.7  \pm     1.7$&$-0.45\pm 0.45$&$                           +0.17 \pm 0.40$\\     
15 &w/c&  10 \, 03 \, 22.0 &  68 \, 45 \, 03.2 &$   12.1 \pm   3.6     $&$        0.8  \pm   1.0      $&$         4.9  \pm    2.2   $&$       6.4  \pm     2.6$&$-0.87\pm 0.26$&$                             -0.06 \pm 0.29$\\  
$^*16$ &w/c&  10 \, 03 \, 23.0 &  68 \, 40 \, 31.7 &$   32.4 \pm   5.8     $&$        3.5  \pm   2.0      $&$         3.9  \pm    2.0   $&$      25.0  \pm     5.1$&$-0.78\pm 0.13$&$                         -0.54 \pm 0.15$\\      
$^*17$ &w/x&  10 \, 03 \, 26.6 &  68 \, 39 \, 22.5 &$    9.4 \pm   3.3     $&$        0.0  \pm   0.0      $&$         1.8  \pm    1.4   $&$       7.9  \pm     3.0$&$-1.00\pm 0.11$&$                         -0.63 \pm 0.26$\\       
$^*18$ &w&  10 \, 03 \, 26.8 &  68 \, 41 \, 29.6 &$    3.8 \pm   2.0     $&$        0.0  \pm   0.0      $&$         0.0  \pm    0.0   $&$       3.9  \pm     2.0$&$-1.00\pm 0.00$&$                           -1.00 \pm  0.00$\\                                    
$^*19$ &w&  10 \, 03 \, 27.2 &  68 \, 41 \, 59.3 &$    5.8 \pm   2.4     $&$        0.0  \pm   0.0      $&$         1.0  \pm    1.0   $&$       4.9  \pm     2.2$&$-1.00\pm 0.13$&$                           -0.66 \pm 0.31$\\       
$^*20$ &w/c&  10 \, 03 \, 29.0 &  68 \, 44 \, 25.3 &$   11.5 \pm   3.5     $&$        0.0  \pm   0.0      $&$         1.9  \pm    1.4   $&$       9.7  \pm     3.2$&$-1.00\pm 0.09$&$                         -0.67 \pm 0.22$\\       
$^*21$ &w/c&  10 \, 03 \, 29.9 &  68 \, 44 \, 22.8 &$   18.5 \pm   4.4     $&$        4.8  \pm   2.2      $&$         3.9  \pm    2.0   $&$       9.7  \pm     3.2$&$ -0.48\pm 0.25$&$                        -0.05\pm 0.24$\\  
22 & w/c& 10 \, 03 \, 30.0 &  68 \, 40 \, 41.8 &$   32.5 \pm   5.8     $&$        4.7  \pm   2.2      $&$         3.9  \pm    2.0   $&$      24.0  \pm     5.0$&$-0.71\pm 0.14$&$                             -0.47 \pm 0.16$\\   
$^*23$ &w/c&  10 \, 03 \, 31.0 &  68 \, 41 \, 54.2 &$   20.3 \pm   4.6     $&$        1.8  \pm   1.4      $&$         1.0  \pm    1.0   $&$      17.6  \pm     4.2$&$-0.82\pm 0.14$&$                         -0.73 \pm 0.16$\\     
$^*24$ &w/c&  10 \, 03 \, 32.9 &  68 \, 46 \, 17.0 &$   31.0 \pm   5.7     $&$        7.8  \pm   2.8      $&$         7.9  \pm    2.8   $&$      15.3  \pm     4.0$&$-0.50\pm 0.19$&$                         +0.01 \pm 0.18$\\ 
25 &w&  10 \, 03 \, 33.4 &  68 \, 40 \, 05.2 &$    3.9 \pm   2.0     $&$        0.0  \pm   0.0      $&$         1.0  \pm    1.0   $&$       2.9  \pm     1.7$&$-1.00\pm 0.22$&$                       -0.49 \pm 0.44$\\     
$^*26$ &w/c&  10 \, 03 \, 41.8 &  68 \, 41 \, 29.7 &$  150.9 \pm  12.3     $&$       15.7  \pm   4.0      $&$        32.8  \pm    5.7   $&$     102.4  \pm    10.1$&$-0.79\pm 0.06$&$                         -0.36 \pm 0.08$\\   
$^*27$ &w&  10 \, 03 \, 43.7 &  68 \, 44 \, 33.5 &$   11.1 \pm   3.5     $&$        0.8  \pm   1.0      $&$         0.0  \pm    0.0   $&$      10.4  \pm     3.3$&$-0.86\pm 0.17$&$                           -0.86 \pm 0.17$\\    
28 &w/c&  10 \, 03 \, 45.9 &  68 \, 41 \, 55.7 &$   21.8 \pm   4.8     $&$        1.7  \pm   1.4      $&$         1.9  \pm    1.4   $&$      18.3  \pm     4.4$&$-0.84\pm 0.14$&$                             -0.67 \pm 0.16$\\  
29 &w/c&  10 \, 03 \, 49.0 &  68 \, 42 \, 56.3 &$   19.5 \pm   4.5     $&$        0.9  \pm   1.0      $&$         3.9  \pm    2.0   $&$      14.7  \pm     3.9$&$-0.91\pm 0.14$&$                             -0.51 \pm 0.20$\\ 
$^*30$ &w/c&  10 \, 03 \, 49.1 &  68 \, 45 \, 51.1 &$   29.7 \pm   5.6     $&$        1.8  \pm   1.4      $&$         4.8  \pm    2.2   $&$      23.1  \pm     4.9$&$-0.88\pm 0.12$&$                         -0.56 \pm 0.16$\\    
$^*31$ &w&  10 \, 03 \, 53.0 &  68 \, 42 \, 22.9 &$   13.7 \pm   3.7     $&$        0.0  \pm   0.0      $&$         3.0  \pm    1.7   $&$      10.8  \pm     3.3$&$-1.00\pm 0.10$&$                           -0.57 \pm 0.22$\\       
32 &w&  10 \, 03 \, 54.6 &  68 \, 41 \, 24.8 &$   15.9 \pm   4.5     $&$        3.3  \pm   2.0      $&$         3.3  \pm    2.0   $&$       9.4  \pm     3.5$&$-0.59\pm 0.27$&$                       -0.18 \pm 0.28$\\    
33 &w& 10 \, 03 \, 55.1 &  68 \, 41 \, 08.6 &$    6.4 \pm   2.6     $&$        0.8  \pm   1.0      $&$         0.0  \pm    0.0   $&$       5.6  \pm     2.5$&$-0.75\pm 0.29$&$                        -0.75 \pm 0.29$\\     
$^*34$ &w/x&  10 \, 03 \, 57.2 &  68 \, 40 \, 44.4 &$   11.2 \pm   3.8     $&$        0.0  \pm   0.0      $&$         2.6  \pm    1.7   $&$       9.9  \pm     3.3$&$-1.00\pm 0.11$&$                         -0.58 \pm 0.24$\\             
$^*35$ &w/c&  10 \, 03 \, 58.5 &  68 \, 43 \, 41.6 &$  265.2 \pm  16.4     $&$       37.3  \pm   6.2      $&$        82.6  \pm    9.1   $&$     145.4  \pm    12.1$&$-0.72\pm 0.06$&$                         -0.10 \pm 0.06$\\  
$^*36$ &w/c&  10 \, 04 \, 02.8 &  68 \, 42 \, 19.4 &$   17.8 \pm   4.7     $&$        0.0  \pm   0.0      $&$         1.5  \pm    1.4   $&$      17.3  \pm     4.5$&$-1.00\pm 0.04$&$                         -0.84 \pm  0.14$\\              
$^*37$ &w/x&  10 \, 04 \, 16.2 &  68 \, 44 \, 35.4 &$   25.6 \pm   5.4     $&$        2.2  \pm   1.7      $&$         3.7  \pm    2.0   $&$      19.7  \pm     4.7$&$-0.83\pm 0.15$&$                         -0.54 \pm 0.18$\\    
$^*38$ &w/x&  10 \, 04 \, 29.4 &  68 \, 43 \, 36.6 &$   60.6 \pm   8.8     $&$        9.0  \pm   3.7      $&$        15.3  \pm    4.2   $&$      36.3  \pm     6.8$&$-0.70\pm 0.14$&$                         -0.20\pm 0.14$\\     
\enddata

\tablenotetext{a}{Right ascension is given in $h~m~s$, Declination in \degr~\arcmin~\arcsec}
\tablenotetext{b}{Foreground star ADS\,7611\,A (binary system with Object No.\,4, HD\,86677)}
\tablenotetext{c}{Foreground star HD\,86677 (BD\,+69\,552)}
\tablenotetext{d}{8\arcsec\ distance to star J100311.93+684148.3 \citep{sak01}}
\label{tab:field}
\end{deluxetable}

\clearpage

\begin{deluxetable}{llrrrrrr}

\tablecolumns{8}
\tabletypesize{\scriptsize}
\rotate
\tablewidth{0pt} 
\tablecaption{Spectral fitting results of the point sources in NGC\,3077.}
\tablehead{
\multicolumn{1}{l}{Parameter}&\multicolumn{1}{l}{Unit}&\multicolumn{1}{c}{S1}&\multicolumn{1}{c}{S2}&\multicolumn{1}{c}{S3}&\multicolumn{1}{c}{S4}&\multicolumn{1}{c}{S5}&\multicolumn{1}{c}{S6}}

\startdata
\multicolumn{1}{l}{RA (J2000)}&&10$^h$03$^m$18\fs8  & 10$^h$03$^m$19\fs1 & 10$^h$03$^m$19\fs1 & 10$^h$03$^m$17\fs8 & 10$^h$03$^m$17\fs9 & 10$^h$03$^m$18\fs3\\
\multicolumn{1}{l}{DEC (J2000)}&&68\degr43\arcmin56\farcs4 & 68\degr44\arcmin01\farcs4 & 68\degr44\arcmin02\farcs3 & 68\degr44\arcmin16\farcs0 & 68\degr43\arcmin57\farcs3 & 68\degr44\arcmin03\farcs8\\

\tableline

Photons&[cts]&$133\pm 12$&$114 \pm 11$&$119 \pm 11$&$37\pm 7$&$17\pm 4$&$17\pm 4$\\*
Count Rate&[10$^{-3}$ cts s$^{-1}$]&$2.50\pm 0.21$&$2.13 \pm 0.21$&$2.23 \pm 0.21$&$0.69\pm 0.13$&$0.32\pm 0.07$&$0.32\pm 0.07$\\

\cutinhead{RS Model}

$N_{H}$& [10$^{21}$\,cm$^{-2}$]   & $9.50_{-1.62}^{+1.47}$ &$13.67_{-2.17}^{+3.76}$ &  $5.38_{-1.10}^{+1.50}$ & $2.78_{-1.28}^{+0.52}$ & $0.98_{-0.88}^{+2.19}$ & $6.01_{-2.04}^{+2.42}$ \\*
$T$& [10$^{6}$\,K]                &$9.10_{-0.85}^{+0.87}$&$102_{-41.2}^{+159}$ & $92.83_{-29.64}^{+84.90}$&$0.65_{-0.01}^{+0.22}$&$7.86_{-1.54}^{+0.66}$&$4.37_{-1.42}^{+1.86}$\\*
Norm.\tablenotemark{a}&  &$3.28_{-0.76}^{+1.00}$ &$2.85_{-0.19}^{+0.82}$ &$2.50_{-0.23}^{+0.47}$ &$304_{-207}^{+64}$ &$0.06_{-0.02}^{+0.11}$ &$0.58_{-0.28}^{+4.96}$ \\*
$F_{X}^{abs}$ &[$10^{-15}$\,erg~cm$^{-2}$~s$^{-1}$]&$10.48_{-4.37}^{+7.57}$&$39.88_{-11.47}^{+16.87}$&$39.55_{-9.60}^{+12.53}$&$2.33_{-1.97}^{+5.22}$&$1.36_{-0.92}^{+4.46}$&$1.40_{-1.25}^{+41.51}$\\*
$F_{X}$ &[$10^{-15}$\,erg~cm$^{-2}$~s$^{-1}$]&$96.58_{-32.17}^{+45.65}$&$63.16_{-9.74}^{+17.25}$&$54.89_{-8.74}^{+11.73}$&$501_{-351}^{+564}$&$2.06_{-0.68}^{+4.10}$&$17.09_{-9.35}^{+181}$\\*
$L_X$ & [10$^{37}$\,erg~s$^{-1}$]&$14.98_{-4.99}^{+7.08}$ &$9.79_{-1.51}^{+2.67}$ &$8.51_{-1.36}^{+1.82}$ &$77.67_{-54.38}^{+87.47}$ &$0.32_{-0.11}^{+0.64}$ &$2.65_{-1.45}^{+28.0}$ \\

\cutinhead{Power Law Model}

$N_{H}$& [10$^{21}$\,cm$^{-2}$]   5& $5.55_{-1.13}^{+1.27}$ & $14.48_{-3.35}^{+5.17}$ & $3.33_{-1.00}^{+1.22}$ & $0.55_{-0.45}^{+0.99}$ & $4.08_{-1.72}^{+3.33}$ & $10.33_{-2.30}^{+3.15}$ \\*
$\gamma$&                       & $3.25_{-0.39}^{+0.43}$ &$1.65_{-0.35}^{+0.48}$ &$1.00_{-0.25}^{+0.26}$ &$5.03_{-0.66}^{+1.44}$ &$4.54_{-1.00}^{+2.13}$ &$8.27_{-0.60}^{+1.73}$ \\*
Amplitude&  [10$^{-6}$]         &$15.57_{-4.16}^{+6.05}$ &$9.09_{-2.28}^{+11.15}$ &$3.77_{-0.79}^{+1.11}$ &$0.32_{-0.11}^{+0.13}$ &$2.36_{-1.03}^{+5.85}$ &$19.08_{-0.85}^{+38.14}$ \\*
$F_{X}^{abs}$ &[$10^{-15}$\,erg~cm$^{-2}$~s$^{-1}$]&$13.65_{-7.27}^{+15.61}$&$41.04_{-27.68}^{+129}$&$58.05_{-28.67}^{+58.63}$&$3.53_{-3.03}^{+76.43}$&$1.58_{-1.32}^{+27.41}$&$1.41_{-1.14}^{+11.60}$\\*
$F_{X}$ &[$10^{-15}$\,erg~cm$^{-2}$~s$^{-1}$]&$124_{-51.7}^{+119}$&$71.56_{-31.45}^{+161}$&$65.14_{-29.14}^{+59.03}$&$14.27_{-9.58}^{+101}$&$62.39_{-49.15}^{+2650}$&$49000_{-37000}^{+116000}$\\*
$L_X$& [10$^{37}$\,erg s$^{-1}$]&$19.25_{-8.01}^{+18.51}$ &$11.10_{-4.88}^{+24.95}$ &$10.10_{-4.52}^{+9.15}$ &$2.21_{-1.49}^{+15.72}$ &$9.67_{-7.62}^{+411}$ &$7570_{-5760}^{+180000}$ \\

\cutinhead{Black Body Model}

$N_{H}$& [10$^{21}$\,cm$^{-2}$]   &$0.98_{-0.70}^{+0.89}$ & $7.26_{-2.21}^{+3.70}$ & $0.58_{-0.48}^{+0.90}$ & $0.38_{-0.38}^{+0.61}$ & $0.08_{-0.08}^{+1.50}$ & $4.36_{-1.63}^{+1.31}$ \\*
$T$& [10$^{6}$\,K]                 &$5.88_{-0.55}^{+0.46}$&$12.04_{-1.57}^{+0.37}$ &$13.62_{-1.38}^{+1.19}$&$0.94_{-0.12}^{+0.11}$&$3.55_{-1.00}^{+0.29}$&$1.81_{-0.19}^{+0.34}$\\*
Amplitude\tablenotemark{b}&        & $2.19_{-0.65}^{+1.24}$ &$0.29_{-0.12}^{+0.23}$ &$0.20_{-0.06}^{+0.09}$ &$2022_{-1820}^{+6020}$ &$1.88_{-0.01}^{+31.08}$ &$234_{-134}^{+287}$ \\*
$F_{X}^{abs}$ &[$10^{-15}$\,erg~cm$^{-2}$~s$^{-1}$]&$12.95_{-7.59}^{+17.37}$&$28.61_{-19.90}^{+60.98}$&$37.49_{-20.49}^{+40.76}$&$2.88_{-2.83}^{+36.92}$&$1.63_{-1.42}^{+37.34}$&$1.38_{-1.20}^{+14.20}$\\*
$F_{X}$ &[$10^{-15}$\,erg~cm$^{-2}$~s$^{-1}$]&$14.92_{-7.83}^{+16.69}$&$34.25_{-22.43}^{+65.87}$&$38.18_{-20.29}^{+40.30}$&$5.92_{-5.63}^{+33.88}$&$1.69_{-1.25}^{+38.78}$&$13.04_{-9.51}^{+47.21}$\\*
$L_X$& [10$^{37}$\,erg s$^{-1}$]&$2.31_{-1.21}^{+2.59}$ &$5.31_{-3.48}^{+10.22}$ &$5.92_{-3.15}^{+6.25}$ &$0.92_{-0.87}^{+05.25}$ &$0.26_{-0.19}^{+6.01}$ &$2.02_{-1.47}^{+7.32}$ \\
\enddata

\tablenotetext{a}{The normalization is given in units of 10$^{-5}\,K$, where $K$ is  
10$^{-14}\,(4\pi D_{A}^{2})^{-1}\,\int n_{e} n_{p} dV$; $D_{A}$
is the angular size distance to the source [cm], $n_{e}$ and $n_{p}$ are the
electron and proton densities [cm$^{-3}$].}
\tablenotetext{b}{The amplitude is given in units of $10^{-5}\,A$, where
$A=2 \pi c^{-2} h^{-3} (R/d)^2 = 9.884 \times 10^{31} (R/d)^2$; $c$ is the speed of light [cm s$^{-1}$], $h$ is Planck's constant [keV\,s], 
$R$ is the radius, and $d$ the distance to the source.}

\label{tab:pointsources}
\end{deluxetable}

\clearpage
\begin{deluxetable}{llrrrrrrr}
\tablecolumns{9}
\tabletypesize{\scriptsize}
\rotate
\tablewidth{0pt} 
\tablecaption{Best fit parameters of spectral RS and MeKaL models to the regions R1 to R7.}
\tablehead{
\multicolumn{1}{l}{Parameter}&\multicolumn{1}{l}{Unit}&\multicolumn{1}{c}{R1 (Total)}&\multicolumn{1}{c}{R2}&\multicolumn{1}{c}{R3}&\multicolumn{1}{c}{R4}&\multicolumn{1}{c}{R5}&\multicolumn{1}{c}{R6}&\multicolumn{1}{c}{R7}}

\startdata
Area& [arcsec$^2$]&8463&268&256&294&341&228&220\\*
Net Counts&[cts]&$729\pm 46$&$142\pm 14$&$104\pm 12$&$43\pm 9$&$58\pm 11$&$71\pm 10$&$58\pm 10$\\*
Net Count Rate& [10$^{-3}$ cts s$^{-1}$]&$13.7\pm 0.4$&$2.7\pm 0.0$&$1.9\pm 0.2$&$0.8\pm 0.2$&$1.1\pm 0.2$&$1.3\pm 0.2$&$1.1\pm 0.2$\\

\cutinhead{RS Best Fit Parameters}

$N_{H}$& [10$^{21}$\,cm$^{-2}$] &$6.31_{-0.32}^{+0.68}$&$6.54_{-1.41}^{+1.00}$&$4.14_{-1.61}^{+2.00}$&$7.67_{-1.99}^{+10.33}$&$5.16_{-1.35}^{+11.40}$&$5.07_{-1.34}^{+1.56}$&$9.41_{-1.01}^{+0.86}$\\*
$T$& [10$^{6}$\,K] &$2.03_{-0.34}^{+0.01}$&$2.02_{-0.28}^{+0.40}$&$2.76_{-0.67}^{+0.80}$&$1.27_{-0.37}^{+1.10}$&$1.80_{-0.03}^{+3.67}$&$2.31_{-0.37}^{+0.55}$&$1.59_{-0.16}^{+0.25}$\\*
Norm.\tablenotemark{a}&&$129_{-6.23}^{+317}$&$29.41_{-21.57}^{+43.42}$&$1.76_{-1.04}^{+11.76}$&$194_{-119}^{+122}$&$9.77_{-8.21}^{+139}$&$4.06_{-2.36}^{+17.26}$&$157_{-136}^{+169}$\\*
$F_{X}^{abs}$& [$10^{-15}$\,erg~cm$^{-2}$~s$^{-1}$]&$31.94_{-22.4}^{+99.7}$&$6.52_{-5.93}^{+55.7}$&$3.33_{-3.11}^{+81.7}$&$1.98_{-1.97}^{+189}$&$2.68_{-2.67}^{+458}$&$3.08_{-2.35}^{+49.7}$&$2.68_{-2.52}^{+20.1}$\\*
$F_{X}$& [$10^{-15}$\,erg~cm$^{-2}$~s$^{-1}$]&$2900_{-641}^{+7190}$&$663_{-513}^{+1140}$&$45.14_{-28.7}^{+319}$&$2300_{-1970}^{+5170}$&$194_{-164}^{+4790}$&$98.00_{-10.1}^{+451}$&$2700_{-2390}^{+3960}$\\*
$L_{X}$& [10$^{37}$\,erg s$^{-1}$]&$450_{-99.4}^{+1110}$&$103_{-79.6}^{+177}$&$7.00_{-4.45}^{+49.6}$&$356_{-306}^{+801}$&$30.05_{-25.4}^{+74.4}$&$15.20_{-1.57}^{+70.0}$&$418_{-371}^{+6114}$\\

\cutinhead{MeKaL Best Fit Parameters}

$N_{H}$& [10$^{21}$\,cm$^{-2}$] &$5.01_{-0.41}^{+0.72}$&$4.61_{-1.38}^{+0.79}$&$4.68_{-1.43}^{+1.43}$&$7.58_{-1.15}^{+0.37}$&$3.94_{-1.35}^{+1.69}$&$0.4_{-0.4}^{+1.13}$&$9.29_{-0.99}^{+0.59}$\\*
$T$& [10$^{6}$\,K] &$2.34_{-0.25}^{+0.08}$&$2.60_{-0.34}^{+0.60}$&$2.48_{-0.46}^{+0.62}$&$1.25_{-0.11}^{+0.23}$&$2.18_{-0.37}^{+0.45}$&$4.88_{-1.05}^{+0.92}$&$1.38_{-0.03}^{+0.29}$\\*
Norm.\tablenotemark{a}&&$46.07_{-29.1}^{+28.9}$&$5.56_{-6.32}^{+6.24}$&$3.47_{-1.65}^{+22.57}$&$253_{-80.2}^{+81.5}$&$2.90_{-1.56}^{+32.06}$&$0.15_{-0.03}^{+0.25}$&$330_{-78.7}^{+79.8}$\\*
$F_{X}^{abs}$ &[$10^{-15}$\,erg~cm$^{-2}$~s$^{-1}$]&$32.82_{-12.4}^{+43.3}$&$6.58_{-5.46}^{+45.5}$&$3.46_{-3.06}^{+82.7}$&$2.00_{-1.39}^{+11.1}$&$2.74_{-2.49}^{+115}$&$3.25_{-2.03}^{+7.98}$&$2.70_{-1.71}^{+9.29}$ \\*
$F_{X}$& [$10^{-15}$\,erg~cm$^{-2}$~s$^{-1}$]&$970_{-69.4}^{+1530}$&$124_{-80.2}^{+219}$&$75.96_{-41.2}^{+528}$&$2510_{-1320}^{+1560}$&$58.24_{-35.5}^{+725}$&$4.16_{-1.13}^{+7.11}$&$3940_{-1910}^{+1330}$ \\*
$L_{X}$& [10$^{37}$\,erg s$^{-1}$]&$150_{-10.8}^{+238}$&$19.25_{-12.4}^{+33.9}$&$11.78_{-6.39}^{+81.8}$&$389_{-205}^{+242}$&$9.03_{-5.50}^{+112}$&$0.65_{-0.18}^{+1.10}$&$611_{-296}^{+206}$  \\
\enddata

\tablenotetext{a}{See footnote a in Table\,\ref{tab:pointsources}}
 \label{tab:regions}
\end{deluxetable}

\clearpage
\begin{deluxetable}{llrrrrrrr}
\tablecolumns{9}
\tabletypesize{\scriptsize}
\rotate
\tablewidth{0pt} 
\tablecaption{Derived parameters of the hot gas from the best RS and MeKaL fits.}

\tablehead{
\multicolumn{1}{l}{Parameter}&\multicolumn{1}{l}{Unit}&\multicolumn{1}{c}{R1 (Total)}&\multicolumn{1}{c}{R2}&\multicolumn{1}{c}{R3}&\multicolumn{1}{c}{R4}&\multicolumn{1}{c}{R5}&\multicolumn{1}{c}{R6}&\multicolumn{1}{c}{R7}\\
\tableline
\multicolumn{9}{c}{Derived Parameters of the Hot Gas\tablenotemark{a}~~~ (RS/MeKaL)}}

\startdata
$d_{eq}$& [pc]&1811&322&315&337&306&297&292\\ 
mean line of sight& [pc]  &1207&215&210&225&204&198&195\\
$n_{e} [\times (\xi f_{v})^{-0.5}]$& [cm$^{-3}$]&0.05/0.03&0.30/0.13&0.08/0.11&0.71/0.82&0.19/0.10&0.13/0.02&0.80/1.16\\ 
$EM [\times (\xi f_{v})^{-1}]$& [cm$^{-6}$\,pc]&3.0/1.1&19.4/3.6&1.3/2.5&113.4/151.3&7.4/2.04&3.4/0.1&124.8/262.4\\ 
$P/k [\times (\xi f_{v})^{-0.5}]$& [10$^{5}$\,K\,cm$^{-3}$]&2.0/1.4&12.1/6.8&4.4/5.5&18.0/20.5&6.8/4.4&6.0/2.0&25.4/32.0\\ 
$M_{hot} [\times \xi^{-0.5} f_{v}^{0.5}]$& [$10^4$\,M$_\sun$]&384.2/230.5&13.0/5.6&3.2/4.4&35.2/40.7&7.0/3.7&4.4/0.7&25.8/37.4\\
$E_{th} [\times \xi^{-0.5} f_{v}^{0.5}]$& [$10^{52}$\,erg]&384.1/265.7&12.9/7.2&4.4/5.4&22.0/25.0&6.2/4.0&5.0/1.6&20.2/25.4\\
$t_{cool} [\times \xi^{-0.5} f_{v}^{0.5}]$& [Myr]&27.0/55.9&4.0/11.8&19.9/14.5&2.0/2.0&6.5/14.0&10.4/77.9&1.5/1.3\\
$\dot{M}_{cool}$& [M$_\sun$\,yr$^{-1}$]& 0.142/0.041&0.033/0.005&0.002/0.003&0.176/0.204&0.011/0.003&0.004/0.0001&0.172/0.288\\
$<v_{hot}>$& [km\,s$^{-1}$]&300/380&320/360&370/350&250/250&300/330&340/480&280/260\\

\cutinhead{Predicted Parameters\tablenotemark{b}}
\multicolumn{2}{l}{Notation of Shells in M98}&\multicolumn{1}{c}{\nodata}&\multicolumn{1}{c}{D}&\multicolumn{1}{c}{A}&\multicolumn{1}{c}{\nodata}&\multicolumn{1}{c}{G}&\multicolumn{1}{c}{B}&\multicolumn{1}{c}{J}\\
\tableline
$n_{amb}$& [cm$^{-3}$]&\multicolumn{1}{c}{\nodata}&\multicolumn{1}{c}{0.45}&\multicolumn{1}{c}{0.60}&\multicolumn{1}{c}{\nodata} &\multicolumn{1}{c}{1.09}&\multicolumn{1}{c}{1.55}&\multicolumn{1}{c}{2.34}\\
$L_{mech}$&[10$^{40}$\,erg~s$^{-1}$]&\multicolumn{1}{c}{\nodata}&\multicolumn{1}{c}{2.0}&\multicolumn{1}{c}{2.6}&\multicolumn{1}{c}{\nodata} &\multicolumn{1}{c}{3.9}&\multicolumn{1}{c}{1.6}&\multicolumn{1}{c}{25.7}\\
$L_X$& [10$^{37}$\,erg~s$^{-1}$]&\multicolumn{1}{c}{\nodata}&\multicolumn{1}{c}{11.3}&\multicolumn{1}{c}{38.1}&\multicolumn{1}{c}{\nodata}&\multicolumn{1}{c}{82.4}&\multicolumn{1}{c}{40.0}&\multicolumn{1}{c}{667.8}\\
\tableline
$v_{exp}^{proj}$& [km~s$^{-1}$]&\multicolumn{1}{c}{\nodata}&\multicolumn{1}{c}{106}&\multicolumn{1}{c}{55}&\multicolumn{1}{c}{\nodata}&\multicolumn{1}{c}{51}&\multicolumn{1}{c}{40}&\multicolumn{1}{c}{67}\\
$t_{shell}$& [Myr]&\multicolumn{1}{c}{\nodata}&\multicolumn{1}{c}{1.7}&\multicolumn{1}{c}{8.6}&\multicolumn{1}{c}{\nodata}&\multicolumn{1}{c}{10}&\multicolumn{1}{c}{9.4}&\multicolumn{1}{c}{8.2}\\

\enddata
\tablenotetext{a}{See Sect.\,\ref{sec:bubbles}.}
\tablenotetext{b}{See Sect.\,\ref{sec:theoX}.}
\label{tab:gas}
\end{deluxetable}

\clearpage
\begin{deluxetable}{llrrr}
\tablecolumns{5}
\tablewidth{0pt} 
\tabletypesize{\scriptsize}
\tablecaption{Best fitting MeKaL models for the probable supernova remnants.}
\tablehead{
\multicolumn{1}{l}{Parameter}&\multicolumn{1}{l}{Unit}&\multicolumn{1}{c}{S1}&\multicolumn{1}{c}{S5}&\multicolumn{1}{c}{S6}\\
\tableline
\multicolumn{5}{c}{MeKaL Model}}

\startdata
$N_{H}$& [10$^{21}$\,cm$^{-2}$]&$9.17_{-1.39}^{+1.22}$&$1.18_{-0.78}^{+1.94}$&$4.22_{-1.65}^{+2.12}$\\ 
$T$& [10$^{6}$\,K]&$9.27_{-1.03}^{+1.13}$&$7.36_{-1.94}^{+1.07}$&$5.56_{-1.51}^{+1.60}$\\
Norm.\tablenotemark{a}&&$3.28_{-0.85}^{+1.05}$&$0.08_{-0.01}^{+0.26}$&$0.27_{-0.10}^{+1.84}$\\   
$F_{X}^{abs}$& [$10^{-15}$\,erg~cm$^{-2}$~s$^{-1}$]&$10.44_{-4.69}^{+8.22}$&$1.37_{-0.88}^{+06.35}$&$1.40_{-1.11}^{+20.84}$\\
$F_{X}$& [$10^{-15}$\,erg~cm$^{-2}$~s$^{-1}$]&$85.75_{-26.31}^{+34.15}$&$2.28_{-0.46}^{+7.20}$&$7.52_{-3.30}^{+52.68}$\\
$L_X$&  [10$^{37}$\,erg~s$^{-1}$]&$13.30_{-4.08}^{+5.30}$&$0.35_{-0.07}^{+1.11}$&$1.17_{-0.51}^{+8.17}$\\   

\enddata
\tablenotetext{a}{See footnote a in Table\,\ref{tab:pointsources}} 
\label{tab:sn_sources}
\end{deluxetable}

\clearpage
\begin{deluxetable}{llccccccc}
\tablecolumns{8}
\tabletypesize{\scriptsize}
\tablewidth{0pt} 
\renewcommand{\arraystretch}{0.9}
\rotate
\tablecaption{Comparison\tablenotemark{1}\ of  NGC\,3077, M\,81, and M\,82.}

\tablehead{
Parameter&Unit&NGC\,3077\tablenotemark{a}&M\,81&M\,81\tablenotemark{b}&M\,81\tablenotemark{b,c}&M\,82&M\,82\tablenotemark{b}&Reference\\
(1)&(2)&(3)&(4)&(5)&(6)&(7)&(8)&(9)}

\startdata
Total $L_{X}^{total}$& [$10^{38}$\,erg~s$^{-1}$]&49/19&$283$&\multicolumn{2}{c}{$97$}&$351$&$196$&1,2,3\\

\tableline

Point sources $L_{X}^{point}$& [$10^{38}$\,erg~s$^{-1}$]&4 (8\%/21\%/27\%) &210 (74\%) &27 (28\%) &88 (91\%)  & 165 (47\%) & 10 (5\%)&1,2,3 \\*
Diffuse emission $L_{X}^{diffuse}$& [$10^{38}$\,erg~s$^{-1}$]&45/15 (92\%/79\%/63\%) &70 (25\%) &70 (72\%)& 9 (9\%)   &186 (53\%) &186 (95\%)&1,2,3 \\

\tableline

Total Mass $M_{tot}$& [$10^{10}$\,$M_\sun$]&1.2&\multicolumn{3}{c}{12}&\multicolumn{2}{c}{3.1}&4,4,4\\*
{\sc Hi} Mass $M_{HI}$& [$10^{8}$\,$M_\sun$]&1.3&\multicolumn{3}{c}{27}&\multicolumn{2}{c}{8.8}&5,6,6\\*
Blue Luminosity\tablenotemark{d} $L_{B}$& [$10^{8}$\,$L_\sun$]&14&\multicolumn{3}{c}{227}&\multicolumn{2}{c}{59}&7,7,7\\*
Infrared Luminosity $L_{IR}$& [$10^{8}$\,$L_\sun$]&3.2&\multicolumn{3}{c}{2.5}&\multicolumn{2}{c}{207}&8,8,8\\*
H$\alpha$ Luminosity $L_{H\alpha}$& [$10^{38}$\,erg~s$^{-1}$]&81&\multicolumn{3}{c}{540}&\multicolumn{2}{c}{4962}&5,9,10\\*
SFR\tablenotemark{e}& [M$_\sun$\,year$^{-1}$]&0.06&\multicolumn{3}{c}{0.4}&\multicolumn{2}{c}{4}& --,--,--\\

\cutinhead{Normalized Luminosities}

$L_{X}^{total}/M_{HI}$& [$10^{30}$\,erg~s$^{-1}$/$M_{\sun}$]&37.7/14.6&10.5&\multicolumn{2}{c}{3.6}&39.9&22.3&\\*
$L_{X}^{point}/M_{HI}$& [$10^{30}$\,erg~s$^{-1}$/$M_{\sun}$]&3.1&7.8&1.0&3.3&18.7&1.1&\\*
$L_{X}^{diffuse}/M_{HI}$& [$10^{30}$\,erg~s$^{-1}$/$M_{\sun}$]&34.6/11.5&2.6&2.6&0.3&21.1&21.1&\\

\tableline

$L_{B}/M_{HI}$& [$L_{\sun}/M_{\sun}$]&10.8&\multicolumn{3}{c}{8.4}&\multicolumn{2}{c}{6.7}&\\*
$L_{IR}/M_{HI}$& [$L_{\sun}/M_{\sun}$]&2.5&\multicolumn{3}{c}{0.09}&\multicolumn{2}{c}{23.5}&\\*
$L_{H\alpha}/M_{HI}$& [$10^{30}$\,erg~s$^{-1}$/$M_{\sun}$]&62.3&\multicolumn{3}{c}{20.0}&\multicolumn{2}{c}{564}&\\*
SFR/$M_{HI}$& [$10^{-8}$\,year$^{-1}$]&0.05&\multicolumn{3}{c}{0.01}&\multicolumn{2}{c}{0.45}&\\

\tableline

$M_{hot gas}/M_{HI}$&&0.030/0.018&\nodata&\nodata&\nodata&\multicolumn{2}{c}{0.15}&\\*
$E_{th}/M_{HI}$& [$10^{46}$\,erg/$M_{\sun}$]&3.0/2.0&\nodata&\nodata&\nodata&\multicolumn{2}{c}{41}&\\*
$\dot{M}_{cool}/M_{HI}$& [$10^{-10}$\,yr$^{-1}$]&10.9/3.1&\nodata&\nodata&\nodata&\multicolumn{2}{c}{2.3}&\\

\enddata
\tablenotetext{1}{For all three galaxies we adopt a common distance of 3.6\,Mpc. All numbers taken from the literature are corrected respectively.}
\tablenotetext{a}{for both models: RS/MeKaL; the relative luminosities given in parenthesis refer to RS/MeKaL/Count Rate}
\tablenotetext{b}{without the strong point--like nuclear source}
\tablenotetext{c}{corrected for unresolved point sources}
\tablenotetext{d}{using the extinction--corrected magnitude $m_{B_{T}}^{0}$ and a solar blue magnitude of 5.50.} 
\tablenotetext{e}{Calculated from the H$\alpha$ luminosities via $SFR=L_{H\alpha}/(1.26\times 10^{41}$\,erg~s$^{-1}$)\,M$_\sun$\,yr$^{-1}$ \citep{ken94}.}

\tablerefs
{(1) this work; (2) \citealt{imm01}; (3) S97; (4) estimated by
assuming an $M/L\approx 9$ by \citealt{bro91}; (5) W02; (6)
\citealt{app81}; (7) \citealt{dvo91}; (8) \citealt{yun99}; (9)
\citealt{gre98}; (10)
\citealt{you88}.}

\label{tab:triplet}
\end{deluxetable}

\end{document}